\documentstyle[emulateapj,apjfonts,psfig]{article}
\newcommand{\be}{\begin{equation}}
\newcommand{\ba}{\begin{eqnarray}}
\newcommand{\ee}{\end{equation}}
\newcommand{\ea}{\end{eqnarray}}  
\newcommand{\etal}{et al.\ }
\def\gtsima{$\; \buildrel > \over \sim \;$}
\def\ltsima{$\; \buildrel < \over \sim \;$}
\def\gsim{\lower.5ex\hbox{\gtsima}}
\def\lsim{\lower.5ex\hbox{\ltsima}}
\def\simgt{\lower.5ex\hbox{\gtsima}}
\def\simlt{\lower.5ex\hbox{\ltsima}}
\def\simpr{\lower.5ex\hbox{\prosima}}
 
\def\CIV{C{\sc ~iv}  }

\def\msun{{M_\odot}}
\def\ie{{\frenchspacing\it i.e. }}
\def\eg{{\frenchspacing\it e.g. }}
\begin{document}
\title{Quasar Feedback: the Missing Link in Structure Formation}

\author{Evan Scannapieco\altaffilmark{1} \& S. Peng Oh\altaffilmark{2}}
\altaffiltext{1}{Kavli Institute for Theoretical Physics,
 Kohn Hall, UC Santa Barbara, Santa Barbara, CA 93106}
\altaffiltext{2}{Department of Physics,
UC Santa Barbara, Santa Barbara, CA 93106}

\begin{abstract}
 
We consider the impact of quasar outflows on structure formation. Such
outflows are potentially more important than galactic winds, which
appear insufficient to produce the level of preheating inferred from
X-ray observations of galaxy clusters.  At late times, energetic
material from the densest objects in the centers of galaxies makes its
way into the intergalactic medium (IGM), impacting structures on many scales, much as supernovae impact structures on many
scales within the interstellar medium.   Using a simple analytical
model for the distribution of quasars with redshift, coupled with a
one-dimensional Sedov-Taylor model for outflows, we are able to make
robust statements about these interactions.  As large regions of the
IGM are heated above a critical entropy of  $S_{\rm crit} \approx 100$
keV cm$^2$, cooling become impossible within them, regardless of
changes in density.  On quasar scales, this has the effect of
inhibiting further formation, resulting in the observed fall-off in
their number densities below $z \approx 2$.  On galaxy scales, quasar
feedback fixes the turn-over scale in the galaxy luminosity function
($L\star$) as the nonlinear scale at the redshift of strong feedback.
The galaxy luminosity function then remains largely fixed after this 
epoch, consistent with recent observations and in contrast to the strong
evolution predicted in more standard  galaxy-formation models.
Finally, strong quasar feedback explains why  the intracluster medium
is observed to have been pre-heated to entropy levels {\em just above}
$S_{\rm crit},$ the minimum excess that would not have been erased by
cooling.  The presence of such outflows is completely consistent with
the observed properties of the Lyman-$\alpha$ forest at $z \sim 2$,
but  is expected to have a substantial and detectable impact on
Compton distortions observed in the microwave background and the
multiphase properties of the ``warm-hot'' ($z=0$) circumgalactic
medium.

\end{abstract}

\keywords{quasars: general --
          galaxies: evolution  -- 
          intergalactic medium -- 
          large-scale structure of the universe --
          cosmology: theory}

\section{Introduction}

Non-gravitational heating is known to have played a central role in
the formation of galaxy clusters.  This is most clearly illustrated
by the observed discrepany in the X-ray Luminosity Temperature
($L_X-T$) relationship of the diffuse intracluster medium (ICM). If
non-gravitational heating was unimportant,  the gas density
distribution would be self-similar, resulting in $L_X \propto T^2$
(Kaiser 1986), but instead the observed slope 
steepens considerably for low-temperature clusters  (\eg David \etal
1993; Arnaud \& Evrard 1999; Helsdon \& Ponman
2000). Non-gravitational heating also manifests itself in the
anomalous hot gas fractions observed  in clusters  (\eg Muanwong
\etal 2002; Tornatore \etal 2003), as well as the low overall level of
the X-ray background (Pen 1999). While the origin of this heating
remains unknown, a wide variety of arguments point to the importance
of active galactic nuclei, and in particular the luminous quasars.
Although high-redshift starburst galaxies are observed to  host
massive supernova-driven outflows (\eg Pettini \etal 2001), it is
unlikely that they are able to generate the $\sim$ 1 keV of energy per gas
particle required to account for cluster preheating (Cavaliere, Menci,
\& Tozzi 1998; Balogh, Babul \& Patton 1999; Kratsov \& Yepes 2000; 
Wu, Fabian, \& Nulsen 2000;
Bialek, Evrard \& Mohr 2001;  Borgani \etal 2001; Brighenti \& Mathews
2001; Tozzi, Scharf, \& Norman 2001; Babul et al 2002).  In fact, 
this level of heating
is so high that it greatly exceeds the energy available  from the
Type-II supernovae resulting from metal-enriched stars, as well as the
additional energy input from pair-production supernovae from a
possible generation of massive primordial stars (\eg Scannapieco,
Schneider, \& Ferrara 2003).  Quasars, on the other hand, represent a
vast reservoir of energy that is largely untapped in current
theoretical modeling.

Radiation from quasars should not be significantly coupled  to the
surrounding gas, except perhaps in the innermost cooling flow regions
of the ICM where Compton heating or cooling could be important (Ciotti \&
Ostriker 1997, 2001). Nevertheless, the  overall bolometric
luminosities of quasars are far above the levels necessary to
effectively preheat the intracluster medium (\eg Inoue \& Sasaki
2001;  Nath \& Roychowdhury 2002).  Furthermore, a large fraction of
quasars are observed to host massive outflows of material, with energies 
comparable to their observed luminosities. 
Thus the light observed to be emanating from quasars at intermediate  
redshifts, while not directly causing gas heating, is likely to be strongly
correlated with outflows that provide a substantial source of
non-gravitational energy.

Like their high-redshift counterparts, these outflows would have
consequences beyond their impact on the intergalactic gas.   Detailed
studies of high-redshift dwarfs indicate that starburst-driven shells
are able to strip gas out of nearby overdense regions that would have 
otherwise later formed stars.  This gas is ejected into space,
leaving behind empty ``dark halos'' of non-baryonic matter
(Scannapieco, Ferrara, \& Broadhurst 2000).   As it is much easier to
strip gas from small objects, such winds naturally exert a feedback
effect on neighboring dwarf galaxies  (Scannapieco, Ferrara, \& 
Madau 2002), causing a tremendous decrease in their numbers and 
slowing the formation of further outflows.

The high entropy of preheated gas will also strongly affect  galaxy
formation, as any level of entropy that can perturb gas density
profiles in clusters will obviously have a major impact on the much
shallower potential wells of galaxies. A moderate increase in entropy
has been seen to reduce angular momentum loss during collapse in
numerical simulations,  (Mo \& Mao 2002; van den Bosch \etal 2003),
and may explain the reduced X-ray luminosities of late-type galaxies (Mo
\& Mao 2002).  Further increases in entropy are able to  cut-off the
supply of cold gas available for galaxy formation, resulting in strong
negative feedback,  which may be important in determining
galaxy morphology  (Oh \& Benson 2003), and which has direct
implications for clusters.

The entropy to which clusters have been preheated 
is not arbitrary, but rather corresponds approximately to the
threshold value for cooling within the age of the universe (Voit \&
Bryan 2001). Indeed, for $S > S_{\rm crit} \sim 100$ keV cm$^2$,  gas
can never cool within a Hubble time regardless of the density to
which it is compressed, and is therefore permanently unavailable for
galaxy formation (Oh \& Benson 2003).  Thus if non-gravitational
heating in clusters had raised the gas to a somewhat lower adiabat,
this energy would have been quickly radiated away, and local X-ray
observations would have left us none the wiser as to its presence.
Yet, if clusters had been heated much {\em above} $S_{\rm crit}$, such
observations would have easily detected this excess. While this
coincidence can only result from fine-tuning in a model that accounts
only for gas heating, it is an unavoidable feature of a model that
accounts for its {\em feedback} on the process of galaxy formation
itself.  As quasar outflows heat the gas around them, they shut off
gas cooling, both in the intracluster medium and in the gas that is
accreting into smaller objects that would have otherwise formed into
galaxies.  In this case, the supply of cold gas available for star
formation and quasar fueling is determined as a result of a
strong feedback equilibrium reached between radiative
cooling and quasar heating.

It is intriguing that the redshifts at which the cores of groups and
clusters were assembled ($z\sim 2-3$) also correspond to the peak of
quasar activity. Oh \& Benson (2003) explored the effects of global
preheating at $z \sim 2-3$ on star-formation, cutting off the supply
of cold gas, but did not consider the spatial dependence of entropy
injection. In fact, more highly biased regions should be heated to
high entropy at earlier epochs, implying that star-formation is
self-limiting in the highest density peaks. A similar self-limiting
mechanism may also operate  in high-redshift star formation ($z>10$),
when the residual entropy from early reionization prevents star
formation in low mass halos (Oh \& Haiman 2003). However, early
preheating due to galactic winds at high redshift $z\sim 10$ would
have little effect on the {\it present-day} galaxy population, since
the residual entropy of the IGM heated at high-redshift is too low
(Benson \& Madau 2003).

While several groups have recently studied the relationship between
the formation history of galaxies and the cosmological evolution of
quasars, none have accounted for the global impact of quasars on the
IGM and its consequences for future galaxy formation.  Kauffmann \&
Haehnelt (2000) incorporated a simple scheme for the growth of
supermassive black holes into a semi-analytic model of galaxy
formation, reproducing the observed luminosity function of quasars in
the context of major mergers.  Ciotti \& Van Albada (2001) used the
observed relationship between the masses of supermassive black holes
and the velocity dispersion of their hosts to study the possibility of
substantial dissipationless merging in the formation of elliptical
galaxies.  Quasar formation regulated by local feedback was studied by
Wyithe \& Loeb (2003; hereafter WL03), as will be discussed in more
detail below.  Di Matteo \etal (2003) studied the metallicity
evolution of quasar host galaxies, finding metallicity gradients and
trends consistent with observations.  Haiman, Ciotti, \& Ostriker
(2003) studied how the duty cycle of quasar activity can be
constrained from the optical quasar luminosity function and the masses
of host galaxies.  Finally, and perhaps most directly related, Granato
\etal (2003) studied the co-evolution of quasars and spheroidal
galaxies, and the impact of quasar-driven winds on the gas {\em
  within} such galaxies.

In this paper  we adopt a more global view, in 
an attempt to understand the features that emerge when
quasar feedback is accounted for in current scenarios of structure
formation.  Using a simple analytical model for the distribution of
quasars with redshift, coupled with a one-dimensional Sedov-Taylor
model for the outflows associated with these sources, we analyze
the impact of quasars on galaxy formation, the structure of
the intergalactic medium, and the further formation of quasars
themselves.  Despite the relative simplicity of our approach, we
are able to make robust statements as to the general features of
strong feedback models that differ from more standard scenarios, as
well as indicate which observational constraints can be used to most 
easily discriminate between the two. 

Throughout this paper we restrict our attention  to a Cold Dark Matter
(CDM) cosmological model with parameters $h=0.7$, $\Omega_m$ = 0.3,
$\Omega_\Lambda$ = 0.7, $\Omega_b = 0.045$, $\sigma_8 = 0.87$, and
$n=1$, where $h$ is the Hubble constant in units of 100 km s$^{-1}$ 
Mpc$^{-1}$, $\Omega_m$, $\Omega_\Lambda$, and $\Omega_b$ are the
total matter, vacuum, and baryonic densities in units of the critical
density, $\sigma_8^2$ is the variance of linear fluctuations on the $8
h^{-1}{\rm Mpc}$ scale, and $n$ is the ``tilt'' of the primordial
power spectrum.  These choices are motivated mainly by measurements
of the cosmic microwave background, the number abundance of galaxy
clusters, and high-redshift supernova distance estimates (\eg Spergel
\etal 2003; Eke \etal 1996; Perlmutter \etal 1999).  Finally, we
make use of the Eisenstein and Hu (1999) fit to the transfer function
in computing the matter power-spectrum.

The structure of this work is as follows. In \S2 we describe a simple
model of the quasar luminosity function (WL03),
which relates the formation rate of quasars to the merger rate of 
collapsed dark-mater halos.  In \S3 we consider the scales of quasar 
outflows, and in \S4 we develop a 
model for the spatial distribution of the resulting high-entropy 
material and its feedback on structure formation. 
In \S 5 we consider the impact of this feedback on the formation
of quasars and galaxies, contrast it with higher-redshift
feedback effects, and examine the impact of quasar outflows 
on the observed properties of the IGM at high and low redshifts.
A discussion is given in \S6.

\section{Quasar Population}

The first step in studying the impact of quasars on structure formation 
is to model their overall distribution.  As discussed by 
Yu \& Lu (2003), any such model is limited by two major 
constraints.  The first of these is the observed number density of
quasars, which has been well-measured out to redshifts $\lsim 6$.
Secondly, as quasars are believed to be powered by 
supermassive black holes (\eg Blandford \& Znajek 1977), 
such models must reproduce the measured black hole mass - bulge velocity 
dispersion $(M_{\rm bh} - \sigma_c)$ relation. 

While a wide variety of models of quasar formation have been suggested
within the CDM paradigm (Efstathiou \& Rees 1988; Haehnelt \& Rees
1993; Kauffmann \& Haehnelt 2000; Nulsen \& Fabian 2000; Monaco,
Salucci, \& Danese 2000; Haiman \& Loeb 2001; Wyithe \& Loeb 2002;
Menci \etal 2003;  Bromley, Somerville, \& Fabian 2003), these can be
roughly divided into two types.  In early studies, quasar formation
rates were tied to the formation rates of dark matter halos, as given
by the rate of change in the  Press \& Schechter (1974, hereafter PS) 
mass function
of dark mater halos.  As both  observational constraints and theoretical
tools improved, such models were widely superseded by approaches that
tied quasar formation to mergers between dark mater halos.
Theoretically, this had the advantage of eliminating poorly defined
``halo formation times,'' as well avoiding corrections
due to the fact that the derivative of the PS
mass function counts both the formation of objects of a given
mass as well as their ``destruction'' by merging into larger objects.
Observationally, on the other hand, there is considerable evidence to
support the idea that quasars are  formed in mergers (\eg Osterbrok
1993).   A large fraction of quasar hosts display a disturbed
morphology (\eg Smith \etal 1986), have nearby companions (\eg Vader
\etal 1987), or even posses  features that can be interpreted as
tidal tails (Stockton \& Ridgway 1991).  Furthermore, the existence of the
tight $M_{\rm bh} - \sigma_c$  correlation suggests that
quasar formation is closely related to bulge formation, which 
itself is thought to be triggered by mergers (\eg Barnes
\& Hernquist 1992).  Finally, we note
that even the merger prescription is not enough to model the low-redshift
quasar population, which exhibits a strong decrease at  $z \lesssim
2$.  This is traditionally interpreted as due to the ever-increasing
gas cooling times within larger halos  (\eg Rees \& Ostriker 1977),
but also arises in a model of strong quasar feedback, as we shall see
below.

As the focus of our investigation is not the detailed modeling of the
processes of quasar formation, but rather the global feedback of this
population on structure formation as a whole, we do not attempt a
survey of these quasar models. Rather we make use of the simple and 
elegant model developed in  WL03 and 
Wyithe \& Loeb (2002), which satisfies both types of constraints in 
the context of major mergers.  Furthermore, as the model is constructed only
to reproduce the distribution of $z \gtrsim 2$ quasars, it contains
no prescription for gas cooling and accretion at low redshifts that
might otherwise obscure our results.

Instead, the authors simply
assume that following a merger, a black hole shines at 
its Eddington luminosity
(as observed by \eg Willott, McLure, \& Jarvis 2003),
and returns  a fraction of this energy to the galactic gas,
eventually disrupting its own fuel source.
In this case a quasar shines for the dynamical time
of the cold gas surrounding the black hole, which is
usually located in a disk with a characteristic radius
$(\lambda/\sqrt{2})r_{\rm vir}\sim0.035r_{\rm vir}$ (Mo, Mao, \&
White 1998).  The 
corresponding dynamical time
is then (\eg Barkana \& Loeb 2001)
\be
t_{\rm dyn}\sim0.035 r_{\rm vir}/v_{\rm c} \approx 5 \times 10^7 
\, {\rm yr} \,\left({1+z}\right)^{-3/2},
\label{eq:tdyn}
\ee
where $\lambda\sim 0.05$ is the spin parameter,
$r_{\rm vir}$ is the virial radius, and $v_{\rm c}$ is the circular
velocity of the galactic halo, which can be written as
\be
v_{\rm c}=140 \, {\rm km~s^{-1}} \, M_{12}^{1/3}
\left({1+z}\right)^{1/2},
\label{eq:vc}
\ee
where $M_{12} \equiv  \left(\frac{M_{\rm
halo}}{10^{12}M_{\odot}}\right)$. Note that the timescale given by eq.\
(\ref{eq:tdyn}) is comparable to the $e$-folding time of the black hole
mass or Salpeter time $t_{S} \sim 2.7 \times 10^{7} (\epsilon_{\rm
rad}/0.06)$yr, where $\epsilon_{\rm rad}$ is the radiative efficiency.

Furthermore, this ``self-regulation'' condition amounts  essentially
to assuming that the total energy output from a quasar is
proportional to the depth of the potential well of the halo in which
it is contained, implying $M_{\rm bh} \propto v_{\rm c}^5$.  Note
however that this  index  is slightly larger than the 4-4.5 inferred
from the local $M_{\rm bh}-\sigma_c$ relation (Merritt \& Ferrarese
2001; Tremaine \etal 2002),  the difference having its origin in the
observation that the $v_{\rm c}-\sigma_c$ relation is shallower than
linear (Ferrarese~2002).  In this model the $M_{\rm bh}-v_{\rm c}$
relation is independent of redshift as observed by Sheilds \etal
(2003), with
\be
M_{\rm bh}= 1.4 \times 10^8 \, M_\odot \, F \, 
\left({v_{\rm c}\over 300 \, {\rm km~s^{-1}}}\right)^5.
\ee
Combining this relation with eq.\ (\ref{eq:vc}) gives
\be
M_{\rm bh}(M_{\rm halo},z) \equiv \epsilon_{\rm bh}(M_{\rm halo},z) 
M_{\rm halo},
\label{eq:eps}
\ee
where $\epsilon_{\rm bh}(M_{\rm halo},z) \equiv  \epsilon_{\rm bh,0} 
M_{12}^{\frac{2}{3}}
(1+z)^\frac{5}{2},$ and $\epsilon_{\rm bh,0} \equiv 3.2 \times 10^{-6} F.$
The only adjustable parameter in this
model is $F$, which, in our notation, was taken to be
$0.6$ in WL03, and is taken to be $1.0$ in our fiducial modeling
below.  Observations (\eg  Ferrarase 2002)
suggest that 
$M_{\rm bh} = M_\odot (1.66 \pm 0.32) \times 10^8 
\left[ \sigma_c/(200\,{\rm km}\,{\rm s}^{-1}) \right]^{4.58 \pm 0.52}$ with
$\log_{10}(v_c/300) =  (0.84 \pm 0.09) \log_{10}(\sigma_c/200 {\rm km/s})
+ (0.0 \pm 0.19)$, which would allow for models with 
$F$ values between 0.15 and 7.0 to fall within 
$1 \sigma$ experimental errors, due primarily to the uncertainty in
the relation between $v_c$ and $\sigma_c$.

\subsection{Optical Quasar Luminosity Function}

Having developed a model for quasar formation during mergers, we
are now able to construct a luminosity function, $\Psi(z,L_B),$ 
which measures the comoving
number density of quasars per unit B-band luminosity  ($L_B$).  
Taking the observed relation of $L_{\rm bol} = 10.4 L_B$
(Elvis 1994) means that  $M_{\rm bh}={L_{\rm Edd,B}}/({5.73\times10^3L_\odot
M_\odot^{-1}}),$  and we can invert
eq.\ (\ref{eq:eps}) to compute $M_{\rm halo}$ as
a function $L_B$,  relating $\Psi(z,M)$ to  the merger rate as
\ba
\Psi(z,L_B)&=& \frac{3}{5\epsilon_{\rm bh}}\frac{t_{\rm dyn}(z)}
               {5.73\times10^3L_\odot M_\odot^{-1}} 
\int_{0.25 M_{\rm halo}}^{0.5
M_{\rm halo}}d\Delta M_{\rm halo} \\
& & \times
\frac{dn_{\rm}}{dM'}(z,M')
\frac{d^2N_{\rm merge}}{d\Delta M_{\rm
halo}dt} (z, \Delta M_{\rm halo},z, M_{\rm halo}), \nonumber
\label{eq:LF}
\ea
where $M' \equiv M_{\rm halo} - \Delta M_{\rm halo}.$
Here the comoving density of halos with mass $M$ is given by the usual
PS prescription
\be
\frac{dn}{dM}(z,M) = 
\frac{\bar{\rho}}{M} \left|\frac{d \sigma^2}{d M}(M) \right|
f[\nu(z),\sigma^2(M)],
\label{eq:abundance}
\ee
where $\bar{\rho}$ is the mean matter density,
$\nu(z) \equiv 1.69 D^{-1}(z)$ is 
the ``collapse threshold'' at
a redshift $z$ with $D(z)$ 
the linear growth factor, $\sigma^2(M)$ is the 
the variance of linear fluctuation within spheres containing a given
mass $M$, and
\be 
f(\nu,\sigma^2) \, d \sigma^2 \equiv
\frac{\nu }{\sqrt{2 \pi} \sigma^{3}} 
\exp\left[-\frac{\nu^2}{2 \sigma^2}\right]\, d \sigma^2,
\label{eq:f1pt}
\ee 
is the mass fraction contained in halos with masses associated
with variances in the range from $\sigma^2$ to $\sigma^2 + d \sigma^2$.
Finally 
\be
\frac{d^2N_{\rm merge}}{d M_p \, dt}
(z_p, M_p, z_f, M_f)
 = \frac{d \nu(z_f)}{d t(z_f)} \frac{\partial}{\partial \nu_f} 
f_2 \left[\nu_p,\sigma^2(M_p),\nu_f,\sigma^2(M_f) \right],
\ee
is the rate of mergers  of
halos with masses between $M_p$ and $M_p + d M_p$ and
formation redshifts $z_p$  
into halos of mass $M_f$ at a merger redshift $z_f$ 
and
\be
f_2(\nu_p,\sigma^2_p,\nu_f,\sigma^2_f) \equiv
\frac{f(\nu_p-\nu_f,\sigma^2_p-\sigma^2_f) f(\nu_f,\sigma^2_f)}
{f(\nu_p,\sigma^2_p)},
\ee
is the conditional probability that an object with a mass $M_p$ at a 
redshift $z_p$ will be found in an object with a mass $M_f$ at
$z_f$ (Lacey \& Cole 1993).

\begin{figure}
\centerline{\psfig{figure=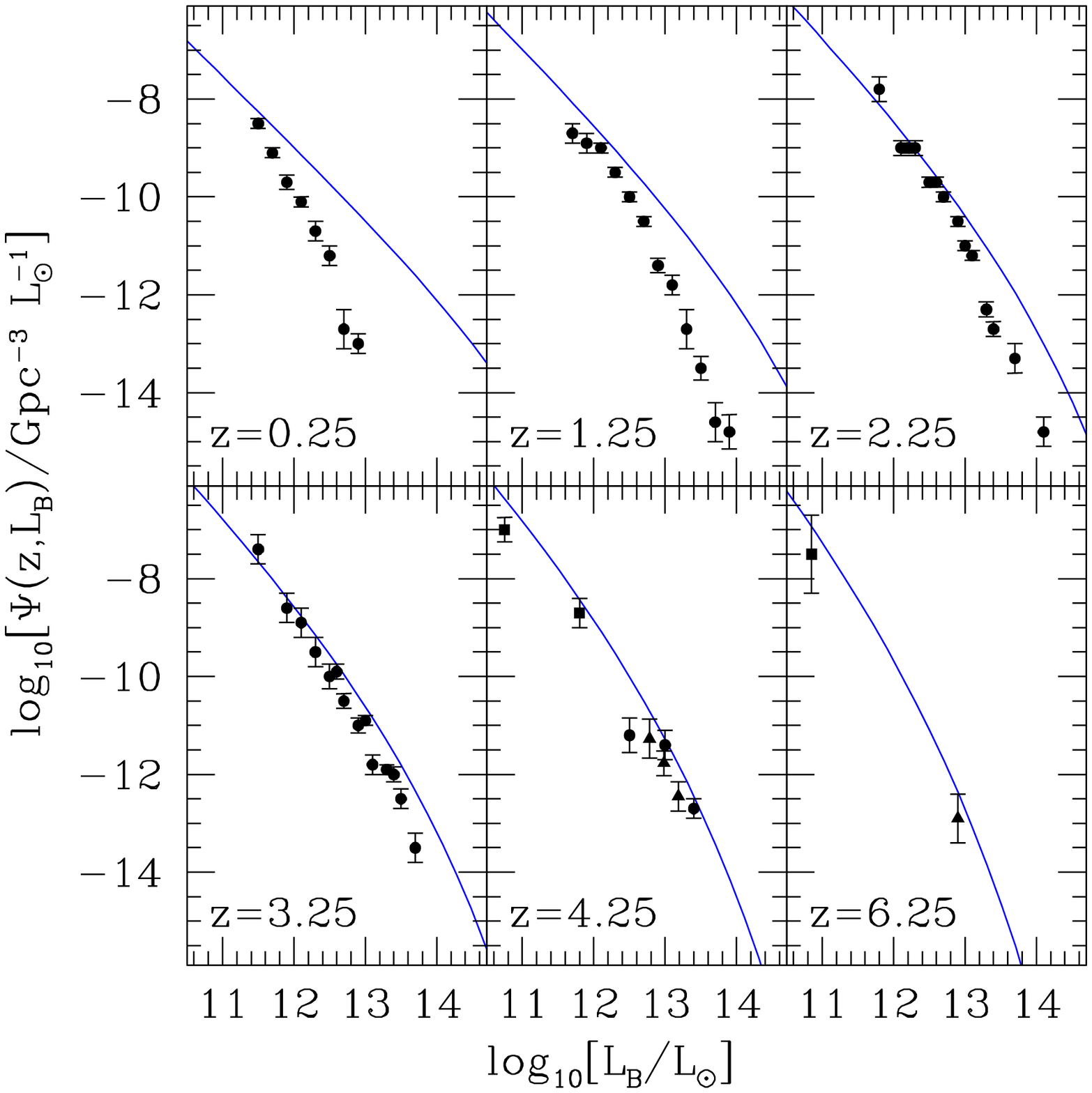,height=11cm}}
%\plotone{f1.eps}
\caption{Evolution of the  B-band quasar luminosity function.
Here the data points are taken from 
Fan (2001, triangles), Pei (1995, circles)
as compiled from Hartwick \& Schade (1990) and 
Warren, Hewett, \& Osmer (1994), and Barger \etal (2003, squares)
as converted from
the X-ray in WL03.  In all panels the solid lines
are the simple WL03 model, with $F = 1.0$.}
\label{fig:lum}
\end{figure}

The resulting luminosity functions are shown in Figure \ref{fig:lum}.
Again we note that the WL03 model is constructed to match the quasar
distribution at high redshifts, and thus assumes that all the gas
within a galactic halo cools on a time much shorter than the Hubble
time.  This is the primary cause of the overestimate of $\Psi$ at the
lower redshifts, as eq.\ (\ref{eq:LF}) includes an additional
contribution from massive halos, which form into groups and clusters
rather than individual galaxies.

\section{Quasar Outflows}

In order to relate $\Psi(z,L_B)$ to the overall level of kinetic
feedback, we must adopt a simple model for the formation and
propagation of quasar outflows.  
Here two major types of objects contribute.  
In radio-loud (RL) quasars, the outflow
takes the form of a collimated jet, which deposits particles into a
cocoon  that expands into the surrounding medium.  These jets, which
are found in  $\sim 10\%$  of all quasars, are observed to have
overall kinetic luminosities ($L_k$) that are correlated with the
bolometric luminosity of the quasars, with $0.05 \lesssim L_k/L_{bol}
\lesssim 1.0$ (Willott \etal 1999).  The major radio-quiet (RQ) sources, on
the other hand, are broad absorption line (BAL) quasars, whose
optical absorption troughs are thought to correspond to outflowing clouds 
with velocities up to  $0.1 c$.  Similar velocities are also observed in
outflowing material detected at X-ray wavelengths (\eg Pounds \etal 2003).
Although these make up approximately
$10\%$ of the observed population of quasars, it is believed that all
quasars host these outflows, with a $10\%$ covering factor (\eg
Weymann 1997). Again observational and theoretical arguments suggest
that the impact of these clouds on the surrounding medium can be
parameterized  with overall kinetic luminosities correlated with the
bolometric luminosities with $0.1 \lesssim L_k/L_{bol} \lesssim 10$
(Furlanetto \& Loeb 2001; Nath \& Roychowdhury 2002).  

Do quasars that host outflows differ widely in their properties from 
the general quasar population?
In the radio-loud case, this relation is a subject of
debate.  A number of earlier studies claimed that, with substantial
overlap, RL quasars tended to be found in Abell 0/1 clusters, while
RQ quasars generally inhabit less-dense regions (\eg
Yee \& Green 1984; Ellingson 1991).  More recent Hubble Space
Telescope studies,
however, have shown that to some degree this is a selection effect,
with a RQ population being found in clusters (\eg Bachall \etal
1997), perhaps in the same relative proportions as RL quasars (McLure
\& Dunlop 2001).

In the BAL case, on the other hand, the primary issue is not if the
spatial distribution of objects is different from the general quasar
population, but rather whether the presence of broad absorption lines
represents a stage in the life cycle of quasars (\eg Briggs \etal
1984), or is simply a geometric effect  (\eg Weymann \etal 1991).   In
this case the geometric interpretation is favored by the similarity
between the optical continuum and line emission of BAL and non-BAL
quasars (Boroson \& Meyers 1992), as well as by the fact that the
sub-millimeter emission of quasar host galaxies  is not related to the
presence of broad absorption lines (Willott \etal 2003).

Finally there is the issue of outflow structure, which is clearly
jet-like in RL quasars, but may later expand roughly spherically
(Furlanetto \& Loeb 2001).  In BAL quasars, outflows are likely to
instead  take the form of axisymmetric winds, accelerated by an unknown
mechanism such as magnetrocentriful forces (\eg Blandford \& Payne
1982) or  radiation pressure from lines (\eg Proga, Stone, \& Kallman
2000).

For a review of the observed relations between RL, BAL, and other quasars,
the reader is referred to Antonucci (1991). For the purposes of this
study however, the important point is only that the populations
of quasars that are observed to host outflows are consistent with
a random subset of all quasars, and that these outflows exhibit a 
variety of structures.  In this case, we adopt the simplifying assumption
that all quasars host outflows of material, with a kinetic
energy input equal to some fraction, $\epsilon_{\rm k} \sim 0.05$, of the total
bolometric energy released over the quasar lifetime (see \S
\ref{section:QLF} for justification of our choice of $\epsilon_{k}$).

Furthermore, the source lifetime $t_{\rm dyn} \sim 10^{7}$yr is
generally shorter than other time scales under consideration. We can
then approximate the energy injection as a point source explosion of
energy $E \approx \epsilon_{\rm k} L_{\rm bol} t_{\rm dyn}$. We shall assume
the bubble evolves adiabatically and ignore energy losses due to
radiative cooling, $p dV$ work against the IGM, and gravity. We also
ignore the Hubble expansion and peculiar velocities (including accretion
infall), and assume that the bubble expands into a medium of uniform
overdensity $\delta_{\rm s}$. The effect of relaxing these assumptions
will be examined later.

\subsection{Outflow Scales}

Under these simplifications, the bubble behaves as a Sedov-Taylor blast
(Sedov 1959) solution, which describes the
adiabatic expansion of a hot sphere of plasma into cold medium.
For our cosmological model, this gives a physical radius of
\be
R_{\rm s} = \xi_{0} \left(\frac{E t^2}{\rho} \right)^{1/5} = 
1.7 \, {\rm Mpc} \,  E_{60}^{1/5} \, \delta_{\rm s}^{-1/5}
\, (1+z)^{-3/5} \, t_{\rm Gyr}^{2/5},
\label{eq:RST}
\ee
where $E_{60}$ is the energy in the hot medium in units of $10^{60}$ ergs,
$t_{\rm Gyr}$ is the expansion time of the bubble in Gigayears,  
the overdensity of the surrounding medium $\delta_{\rm s}$ is defined
such that $\delta_{\rm s} = \rho_b/{\bar \rho_b}$ (rather than
$\rho_b/{\bar \rho_b}-1$)  and
$\xi_{o}=1.17$ for a $\gamma=5/3$ gas (\eg Shu 1992).
Thus the shock velocity is $v_{\rm s} = 2/5 R/t = 1500 \, 
R_{\rm s,Mpc}^{-3/2} \, E_{60}^{1/2} \, \delta_{\rm s}^{-1/2} \, 
(1+z)^{-3/2}$ km s$^{-1}$,
where $R_{\rm s, Mpc} \equiv R_{\rm s} \, {\rm Mpc}^{-1}$.
and, assuming a strong adiabatic shock, the
postshock temperature is $T_{s}=(3 \mu m_{p} v_{s}^{2})/(16 k_{B})$ = 
$13.6 K v_s^2$/(km s$^{-1})^2$.
We therefore obtain the temperature as a function of proper
radius
\begin{equation}
T_{\rm s}= 3.1 \times 10^{7} \, {\rm K} \, E_{60}
\delta_{\rm s}^{-1} \left(1+z \right)^{-3}
R_{\rm s,Mpc}^{-3}.
\end{equation}
The functional form of this equation can be easily understood from $E
\propto \rho R_{\rm s}^{3} (k_{B} T_{\rm s})$= constant, \ie the blast
wave conserves energy. As the postshock density
$\rho_{sh}=(\gamma+1)/(\gamma-1)\rho_{\rm IGM} \delta_{\rm s} = 4 \rho_{IGM}
\delta_{\rm s}$, for $\gamma=5/3$, the postshock entropy can then
be calculated as
\be
S_{\rm s}(r) \equiv \frac{T}{n_{\rm s}^{2/3}} = 1.8 \times 10^4 \, {\rm keV \,
cm^{2}} E_{60}  \delta_{\rm s}^{-5/3} \left( 1+z \right)^{-5} 
R_{\rm s,Mpc}^{-3}.
\label{eq:psS}
\ee
Finally $E_{60}$ can be related to the halo mass and merger 
redshift corresponding to a given quasar, by combining eqs.\ (\ref{eq:tdyn}) 
and (\ref{eq:eps}), yielding
\be
E_{60} = 1.2 \,\epsilon_{\rm k} \, F \, M_{12}^{5/3} \, (1+z).
\label{eq:E60}
\ee

\section{Critical Entropy}

The simple model of entropy injection described above can be easily
extended to capture the global spatial dependence of energy injection
and  cooling, which has not been considered in  previous preheating
models.  The great advantage of considering entropy  is that it is
conserved during adiabatic processes  such as subsonic compression 
and the Hubble expansion, and remains constant as long as
radiative cooling is inefficient.  Moreover, Oh \& Benson (2003)
pointed out that for $\delta (1+z)^{3} > 10$, the cooling time depends
only weakly on density and is always greater than the Hubble time.
Thus, gas that is shocked to entropies greater than a critical value
can never cool,  regardless of the densities to which it is compressed.

This can be most clearly understood if we write the cooling time as a
function of temperature and entropy, rather than temperature and
density. Using $S\equiv T/n^{2/3}$, the isobaric cooling time can be
written as
\begin{equation}
t_{\rm cool}=\frac{(3/2) n k_{B} T}{n_{e}^{2} \Lambda(T)}=S^{3/2}
  \left[ \frac{3}{2} \left( \frac{\mu_{e}}{\mu} \right)^{2}
  \frac{k_{B}}{T^{1/2}\Lambda(T)} \right],
\label{eq:tcool}
\end{equation}
where for a fully ionized gas, $\mu=0.62$, and $\mu_{e}=1.18$. The
quantity in square brackets is a function of temperature only, and
the entropy only changes the overall normalization of the cooling
time.

\begin{figure}
\centerline{\psfig{figure=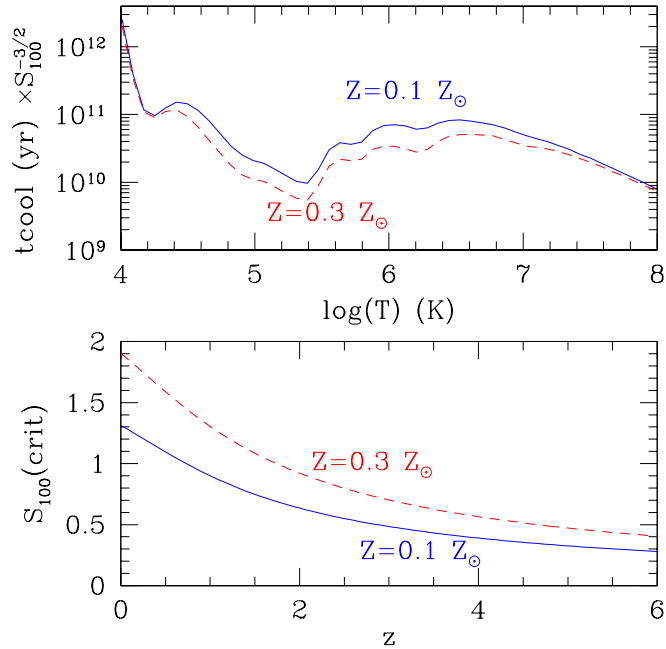,height=9cm}}
%\plotone{f2.eps}
\caption{{\it Top:} The cooling time at constant entropy as a function of
temperature, for $S_{100}=1$. Since $t_{\rm cool} \propto
S_{100}^{3/2}$, changing $S_{100}$ only shifts the curves up or down
the vertical axis. The cooling time has a deep minimum at $T_{\rm
min}=2.3 \times 10^{5}$K. If the cooling time exceeds the Hubble time
at $T_{\rm min}$, it will never be able to cool. {\it Bottom:} 
The critical entropy $S_{\rm crit}$ for which 
$t_{cool}(T_{\rm min}) > t_{H}$, as a function of redshift.}
\label{fig:tcool}
\end{figure}

In the left panel of Figure \ref{fig:tcool}, we plot the cooling time 
at a constant
entropy as a function of temperature for $Z=0.1 Z_{\odot}$ and 
$0.3 Z_{\odot}$, and
$S_{100} \equiv S/(100 \, {\rm keV} \, {\rm cm^2})=
1$. We use fits to the cooling function computed by
Sutherland \& Dopita (1993). Changing $S_{100}$ slides the curves up
and down the vertical axis, but does not change their shape. The
cooling time has a deep minimum at $T_{\rm min}=2.3 \times 10^{5}$K,
where the cooling function peaks. Although this time can be shorter than
$t_{\rm cool}(T_{\rm min})$ at high temperatures, when $\Lambda(T)
\propto T^{1/2}$ due to free-free emission and $t_{\rm cool} \propto
T^{-1}$, this only occurs if $T >
10^{8}$K. As such hot gas can only be retained in the potential
wells of extremely massive
clusters with $T_{\rm vir} > 10$ keV, 
we can safely ignore this regime.

Thus, if $t_{\rm cool}(T_{\rm min}) > t_{H}$, {\it the gas can never
cool in a Hubble time, regardless of any subsequent expansion or
compression}. Since $t_{\rm cool}(T_{\rm min}) \propto S^{3/2}$, this
will occur for all entropies greater than some critical value $S_{\rm
crit}$, given by
\ba
S_{\rm crit} &=& 280 \, {\rm keV \, cm^{2}} \left(1+z\right)^{-1} 
\left[ \frac{E(z)}{(1+z)^{3/2}} \right]^{2/3} \times \nonumber \\ 
& & \qquad \left[
\frac{\Lambda(T_{\rm min})}{ 6.3 \times 10^{-22} \, {\rm erg \, s^{-1}
\, cm^{3}}} \right]^{2/3},
\ea
where $E(z) \equiv \sqrt{\Omega_{m}(1+z)^{3} + \Omega_{\Lambda}}$. We plot
this critical entropy, expressed in terms of $S_{100,{\rm crit}} \equiv
S_{\rm crit} / (100 \, {\rm keV} \, {\rm cm}^2)$ in the right panel
of Figure \ref{fig:tcool}. 

The fact that the entropy floor seen in groups and clusters is greater
than $S_{\rm crit}$ is no coincidence: all lower-entropy gas 
would have dropped out by radiative cooling. Indeed, it was this very 
fact which lead Voit \& Bryan (2001) to suggest that radiative cooling
could cause the observed entropy floor, although
cooling alone could not be wholly responsible, since in the
absence of preheating this would require massive overcooling in groups
(\eg Balogh \etal 2001; Oh \& Benson 2003),
and would result in entropies at large cluster radii that are
much lower than observed (Voit \& Ponman 2003).  The fact that, at 
all radii, the observed
entropy floor is not  far {\em above} $S_{\rm crit},$  however, is
likely to be a manifestation of  strong feedback, as discussed in
detail below.

\subsection{Spatial Distribution}

The presence of a critical entropy suggests that the IGM can be
divided into two fractions, material that has been heated to above
$S_{\rm crit}$ and is therefore unavailable for galaxy formation, and a
cold phase at  $T \sim 10^{4}$K where all memory of preheating has
been erased.  At this point an analogy with the well-known (McKee \&
Ostriker 1977) three-phase model of the interstellar medium (ISM)
within galaxies is irresistible.  In the ISM case, supernova
explosions  result in a three-component medium in which a large
fraction of the  Galactic volume is filled with hot, tenuous gas.  The
remainder of the cold gas is then divided into dense clouds in which
star formation  occurs, and a relatively cool intercloud medium that
accretes onto the clouds that lie outside of hot bubbles.

In the cosmological  case, the $S > S_{\rm crit}$ phase, 
the $T \sim 10^{4}$ phase, and the galaxies themselves play the roles of 
the supernova bubbles,  the cool medium, and the dense clouds, respectively. 
Quasar outflows expand into the
IGM, leaving behind heated regions in which no further structure formation
can occur.  In the cooler zones between these bubbles,
gas accretion continues apace, resulting in the formation of continually 
larger structures.  Finally, the gas within the
galaxies themselves not only condenses to form stars, but also
fuels supermassive black holes, seeding new bubbles of 
$S > S_{\rm crit}$ intergalactic material.

We are interested in the radius $R_{\rm heat}$ 
of the bubble that is heated above the critical entropy, $S > S_{\rm
crit}$. To find this, we consider the sequence of adiabats through
which the gas is shocked and invert eq.\ (\ref{eq:psS}).
This gives a proper radius of
\begin{equation}
R_{\rm heat} = 5.6 \, {\rm Mpc} \, S_{100,\rm
crit}(z)^{-1/3} E_{60}^{1/3} 
\delta_{\rm s}^{-5/9} 
\left( 1+z \right)^{-5/3},
\label{eqn:rcrit}
\end{equation}
\be
\ee
and a corresponding gas mass $M_{\rm b,heat}=(4\pi/3) \, R_{\rm  heat}^{3} 
\rho_{b} \, \delta_{\rm s}$ of
\be
M_{\rm b, heat}(\delta,z,M) =  4.6 \times 10^{12} \, {\rm M_{\odot}} \,
S_{100,\rm crit}(z)^{-1} \,
E_{60} \, \delta_{\rm s}^{-2/3} \left( 1+z \right)^{-2}.
\label{eqn:masscrit}
\ee
If we equate this to the total shocked mass as a function of time,
$M_{s}(t)$, via equation (\ref{eq:RST}), we can find the timescale on
which this entire mass of gas is preheated. It can be quite long for
low-redshift, energetic sources:
\begin{equation}
t_{\rm expand}= 20 \, {\rm Gyr} 
S_{100, \rm crit}(z)^{-5/6} E_{60}^{1/3} \delta^{-8/9} 
(1+z)^{-8/3}.
\label{eqn:texpand}
\end{equation}
Thus, even after the demise of the quasar population at low redshift,
their hot bubbles continue to expand and encompass larger tracts of
the IGM. We discuss this further in section \ref{sect:IGM}.

Unlike the ISM case, the initial positions of
these sources are not determined by baryonic physics or galactic
dynamics, but rather by the underlying dark-matter distribution.
In order to track the evolution of the hot baryonic component,
then, we must combine our outflow modeling with the
formation rate of massive dark matter halos.
Thus, the average number of shells with $S \geq S_{\rm crit}$
impacting a random region of space with an overdensity of 1,
at a redshift $z$ can be computed by 
integrating over the range of masses that contribute to the overall
luminosity function at each redshift.  This gives
\ba 
\left<N_S (z) \right> & = &  \frac{1}{\bar \rho_{b,0}} \int_z^\infty dz' 
\frac{dt'}{dz'} t^{-1}_{\rm dyn} \int^\infty_{M_{40}(z')} 
  d M  \times \nonumber \\
& & \, \, \,  M_{\rm b,min}(z',1,z,M)
\, \frac{d L_B}{d M} \Psi[z',L_B(z',M)],
\label{eq:Ns0}
\ea
where 
${\bar \rho_{b,0}}$ is the mean baryonic density at $z = 0$, 
$M_{\rm 40}(z')$ is the mass corresponding to a circular
velocity of 40 km s$^{-1}$, which approximates the
minimum mass that can form after reionization (\eg Barkana \& Loeb 2001),
and
\be
M_{\rm b,min}(z',\delta_{\rm s},z,M) 
\equiv {\rm min} \left[ M_{\rm b,heat}(1,z',M)
\, ,\,\frac{4 \pi}{3} {\bar \rho_b} \, \delta_{\rm s} \, R_{\rm s}^3(z,z',M) \right],
\label{eq:Mbmin}
\ee
accounts for the possibility that a given outflow may not have expanded to
contain a mass $M_{\rm b,heat}$ by the final redshift $z$.  The factor
of $t_{\rm dyn}^{-1}$ accounts for the fact that the finite duty cycle
of quasars implies that only a fraction $\sim (t_{\rm dyn}/t_{\rm H})$
are visible at any given time; the overall number density of sources
is a factor $\sim (t_{\rm H}/t_{\rm dyn})$ higher than the observed
number density of bright sources.
Assuming that outflows are randomly distributed allows us to simply
relate $\left<N_S(z) \right>$ to
the total mass fraction of gas shocked above $S_{\rm crit}$ as
\be
\Omega_S(z) = 1 - \exp \left[ -\left<N_S(z) \right> \right].
\label{eq:Sfrac}
\ee

As our model associates a mass and
formation redshift with each quasar, it can be easily
generalized to account for the {\em increased} number of
$S \geq S_{\rm crit}$ outflows that impact 
the regions that later form 
into large galaxies and clusters, which is due to the fact that 
quasars tend to be biased towards the densest regions of space.
In this case, the average number of such outflows impacting 
a region  that forms a halo of mass $M_f$ at a redshift $z_f$ is
\ba
\left< N_S(z_f,M_f) \right>
&=&  \frac{1}{\bar \rho_{b,0}} \int_{z_f}^\infty dz' 
\frac{dt'}{dz'} t^{-1}_{\rm dyn} 
\int_{M_{40}(z')}^\infty 
 d M  \times \\
& & \,\,\, 
M_{\rm b, min} [z',\delta_{\rm s}(z',z_f),z_f,M] 
\times \nonumber \\
& & \qquad b(z',M,z_f,M_f)  \,
\frac{d L_B}{d M} \Psi[z',L_B(z',M)], \nonumber
\label{eq:Ns}
\ea
where $b$  is a  now a  ``bias factor'' that accounts for quasar 
clustering and $\delta_{\rm s}(z',z_f)$ 
is the overdensity of the forming object
at the time it encounters an outflow.  

To compute $b$  we employ an analytical formalism that tracks the 
correlated formation of dark matter halos.  
In this model, described in detail in Scannapieco \&
Barkana (2002), objects are associated with peaks in the smoothed linear
density field, in the same manner as the standard PS
approach.  This formalism extends the standard method, however,
using a simple approximation to construct the bivariate mass function
of two perturbations of arbitrary mass and collapse redshift,
initially separated by a fixed comoving distance
(see also Porciani \etal 1998).
From this function we can construct the number density of source
halos of mass $M$ that form at a redshift $z$ at a comoving
distance $r$ from a recipient halo of mass $M_f$ and formation
redshift $z_f$:      
\be
\frac{dn}{d M} (M,z,r|M_f,z_f) =
\frac{d^2 n}{dM dM_f} (M,z,M_f,z_f,r)
\left[\frac{dn}{dM_f} (M_f,z_f)\right]^{-1},
\label{eq:biasnum}
\ee 
where $\frac{dn}{dM_f} (M_f,z_f)$ is the usual PS
mass function and $\frac{d^2 n}{dM dM_f} (M,z,M_f,z_f,r)$ is the
bivariate mass function that gives the product of the differential
number densities at two points separated by an initial comoving
distance $r$, at any two masses and redshifts.  Note that this
expression interpolates smoothly between all standard analytical
limits: reducing, for example, to the standard halo bias expression
described by Mo \& White (1996) in the limit of equal-mass halos at
the same redshift, and reproducing the Lacey \& Cole (1993) progenitor
distribution in the limit of different-mass halos at the same position
at different redshifts.  Note also that in adopting this definition we
are effectively working in Lagrangian space, such that $r$ is the {\em
initial} comoving distance between the perturbations.  Fortunately as
shock propagation is likely to be more closely dependent on the column depth
of material between the source and the recipient than on their
physical separation, this natural coordinate system is more
appropriate for this problem than the usual Eulerian one.  
We then define the bias factor as
\be
b(z,M,z_f,M_f)  = \frac{dn}{d M} (z,M,r|z_f,M_f) \left[ \frac{dn}{d M}(z,M)
\right]^{-1},
\ee
which is simply the number density of mass $M$, redshift $z$
halos near the final object, divided by the average density of
such halos.  Finally for the effective radius, one works in Lagrangian
comoving coordinates, such that $r$ is the virial radius plus the
radius which would contain $M_{\rm b, min}$ in a mean-density medium, 
$r = \left(\frac{3}{4 \pi {\bar \rho_{b,0}}} 
     \, M_{\rm b,min} \right)^{1/3}+r_{\rm vir}(M)$.
Note that this calculation of bias, while carried out within
our merger formalism, depends only on the mass of the quasar
host galaxy, the turn-on redshift for the quasar, and the
mass and formation redshift of the final object.  As these
are the same quantities that arise in an
approach based instead on accretion, one would expect the
relationship between $\left<N_s(z_f,M_f) \right>$
and $\left< N_S(M) \right>$ to be similar in any such model.
This means that the rough features
derived below are likely to persist, regardless of the detailed
physics of quasar generation.

To approximate $\delta_{\rm s}$, we appeal to the standard top-hat
collapse model (see \eg Peebles 1980), which relates the
true overdensity to the ``linear'' overdensity, $\delta_L$
expressing both quantities parametrically
in terms of a collapse parameter, $\theta,$ as
\be
\delta_{\rm s}(z,z_f) = \frac{9}{2} \frac{(\theta - {\rm sin} \, \theta )^2}
{(1 - {\rm cos} \, \theta)^3},
\label{eq:theta1}
\ee
and
\be
\delta_L(z,z_f) = 1+ \frac{3}{5}\left(\frac{3}{4}\right)^{2/3} 
(\theta - {\rm sin} \, \theta)^{2/3} = 1.69 \frac{D(z)}{D(z_f)},
\label{eq:theta2}
\ee
where $D(z)$ is again the linear growth factor.

Finally, we relate $\left< N(z_f,M_f) \right>$
to $f_S(z_f,M_f),$ the fraction of objects 
with collapse redshifts $z_f$ and mass $M_f$
whose formation is suppressed by preheating by quasar outflows. 
As our formalism
can only account for the correlations between two halos, we must adopt
an approximation for the fraction of objects that are affected by
multiple  outflows.  The simplest approach is to 
assume that while outflows only impact a fraction of the overall
cosmological volume, their arrangement {\em within that fraction} is
completely uncorrelated (as in Scannapieco, Schneider, 
\& Ferrara 2003).  In this approximation
$f_S (z_f,M_f)$ and 
$\left <N_S(z_f,M_f)\right>$ share the
same relation as $\Omega_S(z)$ and $\left< N_S(z) \right>$ in
the average case, namely
\be 
f_S(z_f,M_f) = 1 - \exp[- \left< N_S(z_f,M_f) \right>].
\label{eq:SfracM}
\ee
We will adopt this {\em Ansatz} throughout this paper.

\subsection{Theoretical Uncertainties}

Before we apply the formalism developed above to derive the features
expected in a model of strong feedback, we first pause to assess the
uncertainties involved in our approach.  
Primary amongst these is our use of
an adiabatic Sedov-Taylor solution for a point source  explosion in a
uniform medium.  Here we note that 
\be 
t_{\rm heat} \approx
\frac{R_{\rm heat}}{v_{\rm sh}} = 4.8 \times 10^{10} \, {\rm yr} \,
S_{100,\rm crit}^{-5/6} E_{60}^{1/3} \delta_{\rm s}^{-8/9} \left( 1+z
\right)^{-8/3}, 
\ee 
which is larger than our source lifetime of
$t_{o}\sim 10^{7}$yr (as in equation \ref{eq:tdyn}) justifying our
assumption of instantaneous energy injection.

Secondly, we ignored radiative losses. Since we are only interested in
the early stages of the blast wave, when gas is shocked to high entropies
($S > S_{\rm crit}$) and can never cool, this is justified; $R_{\rm heat}$
is by definition the radius at which the cooling time of the gas is
equal to the Hubble time. Beyond $R_{\rm heat}$, radiative cooling
indeed becomes important and the evolution of the blast wave is no
longer adiabatic.

Note that Compton cooling will dominate over collisional cooling at
high redshifts ($z>7$); however, we do not consider such high redshifts
in our calculations: both the high IGM densities and the effectiveness
of Compton cooling mean that the residual entropy is low. We also
ignored the deceleration of the shock due to the $p \, dV$ work it does
on the IGM, which is justified as long as $\rho v_{\rm sh}^{2} \gg
P_{\rm IGM}$, or $T_{\rm shock} \gg T_{\rm IGM} \sim 10^{4}$K, which is
certainly the case. Gravitational deceleration is unimportant as long
as the work done in sweeping up the shell out of the potential well
$W=0.5 \int \rho_{g} \Phi d^{3}x= G \pi M_{\rm halo} \delta \rho_{\rm
IGM} R_{\rm sh}^{2}$ is smaller than the energy of the explosion,
which it is
\be
\frac{W(S>S_{\rm crit})}{E}= 4.5 \times 10^{-2} \, S_{100,\rm
crit}^{-2/3} E_{60}^{ -1/3} 
\delta_{\rm s}^{-1/9} \left(1+z\right)^{-1/3} M_{12}.
\ee
Note that since $E \propto M_{12}^{5/3}$, $W/E \propto M_{12}^{1/6}$, so the dependence on halo mass is very weak. Finally, ignoring Hubble expansion is valid as long as $v_{\rm sh} \gg
H(z) R_{\rm sh}$, which is only marginally valid:
\begin{equation}
\frac{v_{\rm sh}}{H(z) R_{\rm heat}} \approx 
\frac{t_{\rm heat}}{t_{H}} = 3.5 S_{100,\rm crit}^{-5/6}
E_{60}^{1/3} \delta_{\rm s}^{-8/9} \left( 1+z \right)^{-7/6}.
\end{equation}
Our model is therefore conservative in that we underestimate 
the extra kinetic energy associated with the Hubble expansion. 

In order to use the Sedov-Taylor solution, we have assumed a point
source explosion in a homogeneous medium, characterized by some mean
overdensity $\delta$. In fact, the density profile is more likely to
be that of a power law; in addition the medium is not static but is
being accreted by the halo. It might seem therefore that the dynamics
of the explosion are no longer accurately described. In fact, Barkana
\& Loeb (2001) and Furlanetto \& Loeb (2001) have constructed
numerical models of outflows and solved the coupled system of
differential equations to describe the outflow dynamics, modeling
both the radial dependence of the density profile as well as accretion
infall. They find remarkable agreement with a simple Sedov-Taylor type
model such as that used here, which ignores the density profile and
infall, as well as cooling, external pressure of the IGM, and the
gravitational potential of the host halo (see Fig.\ 1 of Furlanetto \&
Loeb (2001), and associated discussion). This can be understood, since
in all cases $R_{\rm heat} \gg R_{\rm vir}$, and the swept-up mass is
dominated by that accumulated at large radii, when the density is
close to the cosmological mean density.  It is therefore largely
independent of the details of the infall region.

We have completely bypassed the issue of how the quasar wind escapes from
the host galaxy. Such issues are beyond the province of our
semi-analytic model and can only be studied with hydrodynamic
simulations. To our knowledge, there has been no detailed numerical
study of quasar winds, although there is a significant body of literature
on starburst winds (e.g., Strickland \& Stevens 1999; MacLow \&
Ferrara 1999; Mori, Ferrara \& Madau 2002; Fujita \etal
2003). Simulations of a centrally concentrated nuclear starburst most
closely approximate quasar winds and generically find that
starbursts punch a hole through the ISM of the galaxy ("blow-out"),
venting most of the energy of the wind (with little radiative losses)
while entraining little cold gas, which remains in the galaxy. Some
fraction (few $\times 10\%$) of the kinetic energy of the outflow can
be radiated in internal, inward propagating shocks if there are
significant off-nuclear supernova explosions (Mori, Ferrara \& Madau
2002), but the radiated energy diminishes for a single, central energy
source. 

In fact, a single source, rather than smoothly distributed
energy injection, is the most efficient means of quickly producing
blow-away (Strickland \& Stevens 1999). The coupling between the ISM
and the wind diminishes rapidly for more energetic bursts, as blow-out
becomes more efficient; only for very small dwarfs $M_{g} \le 10^{6}
\, M_{\odot}$ is a large fraction of the ISM entrained in the outflow
(MacLow \& Ferrara 1999). Blow-out would likely be particularly
efficient for an initially bi-polar outflow such as an quasar wind. Thus,
while an quasar wind has not been directly simulated, it certainly lies
in the region of parameter space where energy should be efficiently
delivered to the IGM. In any case, our model is only applicable once
the wind has propagated out of the host galaxy: all uncertainty in
energy loss within the host galaxy is encapsulated in the free
parameter $\epsilon_{\rm k}$. For simplicity, we have assumed
$\epsilon_{\rm k}$ to be independent of galaxy mass. We emphasize that
our model only taps a very small fraction, $\epsilon_{\rm k} \sim
5\%$, of the total available kinetic luminosity, and so preheating can
still be important even if an appreciable fraction of energy is
radiated away in the ISM of the host.

We approximate the metallicity of the IGM as $0.1 Z_\odot$, consistent
with the level of  pre-enrichment necessary to explain the lack of
low-metallicity G-dwarf stars in the solar neighborhood (\eg Ostriker
and Thuan 1975), as well as in other large nearby galaxies (Thomas,
Greggio, \& Bender 1999). This should also be representative of the
metallicity of regions surrounding quasar outflows at $z=2$, when
preheating by outflows peaks. Such regions now constitute the
intracluster medium, which are observed to have metallicity $Z \approx
0.3 Z_{\odot}$; if $\sim 30\%$ of all present-day stars have formed by
$z\sim 2$, as indicated by the Madau plot, then $Z(z=2) \sim 0.1
Z_{\odot}$. However, assuming a higher metallicity does not greatly alter
$S_{\rm crit}$ (see Fig \ref{fig:tcool}).

The greatest uncertainty is $\epsilon_{k} \equiv L_{k}/L_{\rm Bol}$,
the fraction of the total bolometric luminosity (assumed to be the
Eddington luminosity) which appears as kinetic luminosity. Most
observations suggest that $L_{\rm k} \sim L_{\rm B}$ for both
radio-loud (Willott et al 1999) and BAL quasars (Furlanetto \& Loeb
2001), albeit with at least an order of magnitude uncertainty. Since
$L_{B} \sim 0.1 L_{\rm Bol}$ (Elvis et al 1994), $\epsilon_{k} \sim
0.1$ is reasonably motivated from observations. However, the
uncertainty is so large that we shall simply take $\epsilon_{k}$
to be a free parameter, which we use to match the observed quasar
luminosity function, as described in the following section.

\section{Implications for Structure Formation}

\subsection{Global Features}

From eqs.\ (\ref{eq:Ns}) and (\ref{eq:SfracM}), we can construct
the fraction of galaxies whose formation is suppressed given an
arbitrary luminosity function $\Psi[z,L_{\rm B}(z,M)].$  Yet our 
model is not complete without 
correcting $\Psi$ for the corresponding
suppression in quasar formation.  Naively,
one might consider convolving the luminosity function with the fraction
of objects affected by heating at a given redshift, replacing
$\Psi[z,L_{\rm B}(z,M)]$ with
$[1-f_S(z,M)] \times \Psi[z,L_{\rm B}(z,M)].$
Yet this approximation would result in a severe underestimate, as in reality
two galaxies of mass $M/2$ that formed at early times can still merge 
to ignite a quasar at $z$, even if the formation of additional 
galaxies of mass $M$ is largely excluded at this redshift.

A better approach relies on the fact, as we shall see below, that
$\left< N_S(z,M) \right>$ is in general monotonically increasing
with time.  Thus we define an
``exclusion redshift,'' $z_{\rm ex}(M)$ such that all $M$ mass 
galaxies forming by $z_{\rm ex}(M)$ contribute to the quasar 
luminosity function, while later forming galaxies do not, with 
$z_{\rm ex}$ defined implicitly as
\be
\left<N_S[z_{\rm ex}(M),M] \right> = 1.
\ee
With this definition, we then impose that for all halo masses and
redshifts such that $z \le z_{\rm ex}(M_{\rm halo}/2)$, eq.\
(\ref{eq:LF}), gas can no longer cool and the supply of cold gas is
shut off. Thus, only halos that formed at $z > z_{\rm ex}$ will
contain cold gas to fuel quasars, and only their mergers need be
considered. These progenitor galaxies will in
general reside in a much larger host halo (such as a group or
cluster), whose mass greatly exceeds the sum of the merging 
galaxy masses.
Thus eq.\ (\ref{eq:LF}) must be rewritten to include an
additional integral over possible final masses,
$M_{\rm big}$ of the final object.  Defining $M'_{\rm big} =
M_{\rm big} - \Delta M_{\rm halo}$ this integral can be written as
\ba
\Psi(z,L_B) &=& \frac{3}{5\epsilon_{\rm bh}}\frac{t_{\rm dyn}(z)}
               {5.73\times10^3L_\odot M_\odot^{-1}} 
\int_{0.25 M_{\rm halo}}^{0.5
M_{\rm halo}} d\Delta M_{\rm halo} \times \nonumber
\label{eq:LF2} \\
& & \, \, \frac{dn}{dM'}(z_{\rm ex},M') \int_{M'}^{\infty} d M'_{\rm big} \,
 \frac{d \sigma^{2'}_{\rm big'}}{d M'_{\rm big}} \times \nonumber \\
& & \qquad
f_2 \left[\nu(z_{\rm ex}), \sigma^2(M'),\nu(z),\sigma^{2'}_{\rm big} \right]
\times \nonumber \\ & & \qquad 
\frac{d^2N_{\rm merge}}{d\Delta M_{\rm halo}dt} 
(z_{\rm ex},\Delta M_{\rm halo},z, M_{\rm big}). 
\ea
Note that here we assume that the galaxies contained in each sub-halo
merge instantaneously as they fall into the larger halo $M_{\rm
halo}$, and do not attempt to account for dynamical friction or tidal
stripping within this region (\eg Kauffmann, White, \& Guiderdoni
1993; Cole \etal 1994; Somerville \& Primack 1999)  as these are
beyond the scope of our simple model and would only obscure the
processes in which we are interested.  Note also that in the limit in
which $z_c \rightarrow z$, $f_2$ approaches
$\delta[\sigma^2(M')-\sigma^{2'}_{\rm big}]$ and thus eq.\ (\ref{eq:LF2})
reduces to eq.\ (\ref{eq:LF}).

We then adopt an iterative approach to obtain $\Psi(z,L_B)$ and
$\left<N_S(z,M) \right>$ as a function of our feedback
parameter $\epsilon_{\rm k}$: first computing $\left<N_S(z,M) \right>$
from the simple WL03 model, applying eq.\ (\ref{eq:LF2}) to 
correct the luminosity function, computing $\left<N_S(z,M) \right>$ 
from this corrected function, and so on. In practice this 
convergence is very quick, arriving at a self-consistent solution
within two cycles, although we iterate five times in the
results presented below.

In Figure \ref{fig:Mfinal} we show the $\left<N_S(z_f,M_f) \right>$
values that result from models with $F=1$ and three different values
of $\epsilon_{\rm k}$.  Here we choose $\epsilon_{\rm k} = 0.05$ as our fiducial
model, simply because it provides the best fit to the quasar
luminosity function, as discussed in \S5.2.     We then double and
halve this value to examine the  impact of varying the feedback
efficiency on our modeling, resulting in the  $\epsilon_{\rm k}= 0.025$ and
$\epsilon_{\rm k} = 0.10$ models depicted in the lower panels.  In all cases
$\left<N_S(z_f,M_f) \right>$ is a monotonically decreasing function of
redshift, as assumed.

While in a self-similar, hierarchical model, preheating by progenitors
would be independent of the final host halo mass scale, three features
of quasar outflows break this self-similarity.  As energy injection is
a non-linear function of halo mass, $E \propto M_{\rm bh} \propto
M_{\rm halo}^{5/3},$  feedback is a strongly increasing function of
mass. 
Secondly, the presence of a critical entropy imposes a scale
into the problem,  as the number of
supercritical outflows is self limiting, and thus does not greatly
exceed one even for the largest objects.  
Finally, high mass objects
are more strongly clustered, and therefore more likely to collapse out
of preheated gas than low mass objects. The effect is similar to
well-known fact that gas metallicity has a strong density dependence,
and metals are likely to be clustered around the highest density
peaks. Metals are likely to be a good tracer of gas entropy, and the
two should exhibit similar spatial bias.

Thus for larger masses halos with $M_f \gtrsim 10^{13} \msun$, 
$\left<N(z_f,M_f) \right>$
is a relatively weak function of mass.  This is because as
$\left<N(z_f,M_f) \right> \geq 1,$ the further formation of galaxies
is strongly suppressed, and these objects are heated primarily by
high-redshift $M \ll M_f$ progenitors.  In this limit the bias
is well approximated by the Lacey \& Cole (1993)
progenitor distribution
\be
b (M,z,M_f,z_f) =
\frac{f[\nu(z)-\nu(z_f),\sigma^2(M)-\sigma^2(M_f)]}
     {f[\nu(z),\sigma^2(M)]},
\label{eq:lacey}
\ee 
where $\nu(z)$ and $\nu(z_f)$ are the ``collapse thresholds'' and
$\sigma^2(M)$ and $\sigma^2(M_f)$ are the variances
associated with the progenitor and final objects, respectively.
As $M \ll M_f$, $\sigma^2(M) \gg \sigma^2(M_f)$, and therefore eq.\ 
(\ref{eq:lacey})
is nearly independent of the $M_f$.  In other words, the 
number density  of $M \lesssim 10^{12} \msun$  progenitors is roughly
equivalent for groups and massive clusters that form at the
same redshift, and thus $\left<N(z_f,M_f) \right>$ is
largely equal for these objects.

Similarly, the impact of quasar heating on low-mass, $M\lesssim 10^{13} \msun$
objects has a very different dependency on the feedback parameter
$\epsilon_{\rm k}$ than at higher mass scales.  While  $\left<N(0,10^{10}
\msun) \right>$  changes from $0.05$ to  $0.3$ as $\epsilon_{\rm k}$ goes
from $0.025$ to $0.10$,  $\left<N(0, M \gtrsim 10^{13} \msun) \right>$
changes only from 2 to 4 for the same $\epsilon_{\rm k}$ values.  Again
this is a manifestation of feedback effects, which are almost
nonexistent in the $10^{10} \msun$ case, but  greatly reduce the
number of outflows generated by large galaxies.  In fact,
the true dependence of   $\left<N(0, M \gtrsim 10^{13} \msun) \right>$
on the feedback parameter is likely to be less, as eq.\
(\ref{eq:LF2}) applies the same correction to the luminosity function,
regardless of the overall environment.  In reality, $\Psi(z,L_B)$
should be somewhat more suppressed in clusters than the field, leading
to an even weaker dependence of cluster preheating on $\epsilon_{\rm k}$.

\begin{figure}
\centerline{\psfig{figure=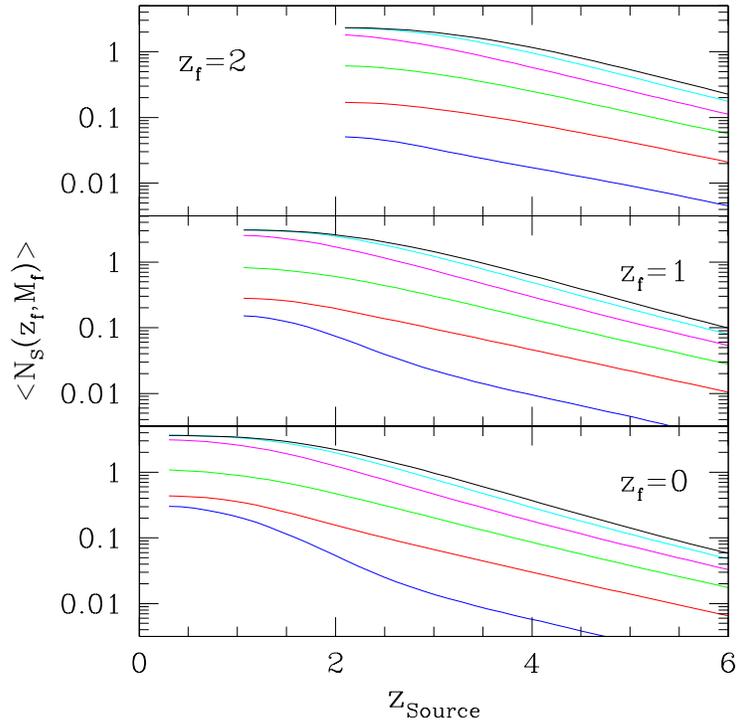,height=10cm}}
%\plotone{f4.eps}
\caption{Preheating of objects in our fiducial  $F=1$, $\epsilon_{\rm k} =
0.05$ model.  Here each panel gives the contribution  to
$\left<N_S(z_f,M_f) \right>$ from sources with redshifts $\ge z_{\rm
source}$.  Panels are labeled by their $z_f$ values, and in all panels
the curves represent masses (from top to bottom) of $10^{15} \msun$,
$10^{14} \msun$, $10^{13} \msun$, $10^{12} \msun$, $10^{11} \msun$,
and $10^{10} \msun$.}
\label{fig:Msource}
\end{figure}

To explore these dependencies further, in Figure \ref{fig:Msource} we plot
the contribution to $\left<N_S(z_f,M_f) \right>$ from sources above
some higher redshift $z_s$, in our fiducial model.
Here we see that in the higher-$z_f$, lower-mass objects,
the impact of quasar halos is extremely weak at all times.
One the other hand, in the $z_f = 0$, lower-mass objects,
a substantial 
upturn in the contribution from sources at lower redshifts
is seen, particularly in the $10^{10} \msun$ case.
This is due to the ability of $S \geq S_{\rm crit}$ 
winds from larger objects to move into
underdense regions if given sufficient time to expand.
Finally, at higher masses, a sizeable contribution to 
$\left<N_S(z_f,M_f) \right>$ is observed at all redshifts,
arising from progenitor objects.

From the values shown in Figure \ref{fig:Mfinal}, we can directly
calculate $z_{\rm ex}(M)$ as well as the fraction of suppressed
galaxies, as computed from eq.\ (\ref{eq:SfracM}).  These are shown in
Figure \ref{fig:zex}, again for three representative $\epsilon_{\rm k}$
values.  This plot provides a different perspective on the trends seen
above.  In our fiducial model, because $f_S(z_f,M \lesssim 10^{11} \msun) \leq
0.25$ at all redshifts, low-mass galaxy formation continues
largely unimpeded to this day.  In the $10^{12} \msun$ to $10^{13}
\msun$ mass range, however, a steep rise in $z_{\rm ex}$ take place as
feedback becomes more effective.  Finally, at masses $\gtrsim
10^{13.5} \msun,$ $z_{\rm ex}$ values are largely independent of mass,
and strong feedback occurs in all cases.  These trends persist in the
$\epsilon_{\rm k} = 0.025$ and $\epsilon_{\rm k} = 0.10$ models although the
overall values shift somewhat.  In general, raising the feedback
efficiency causes the transition region in $z_{\rm ex}$ to move to
smaller masses, while pushing the high-mass $z_{\rm ex}$ plateau to
higher redshifts.

\subsection{Entropy and the Quasar Luminosity Function}
\label{section:QLF}

\begin{figure}
\centerline{\psfig{figure=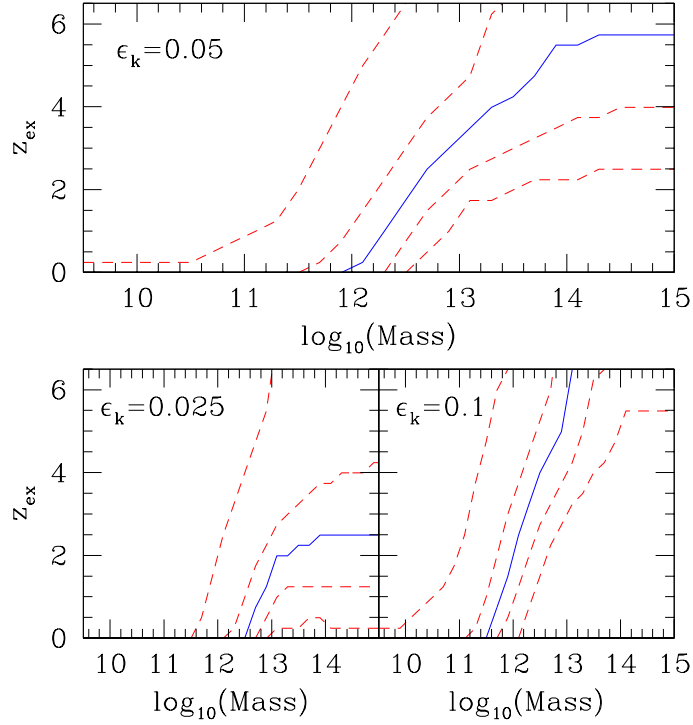,height=9.5cm}}
%\plotone{f5.eps}
\caption{``Exclusion redshifts'' for various models of quasar heating.
In each panel the solid lines show $z_{\rm ex}$ such that $\left< 
N_S(z_{\rm ex},M) \right> = 1,$ while the short-dashed lines correspond
(from top to bottom) to $z$ values such that $f_S(z,M) = 1 - \exp \left[
- \left< N_S(z,M) \right> \right] =$ 0.25, 0.5, 0.75, and 0.9.
The panels are labeled by their assumed $\epsilon_{\rm k}$ values, and in
all cases $F=1$.} 
\label{fig:zex}
\end{figure}

\begin{figure}
\centerline{\psfig{figure=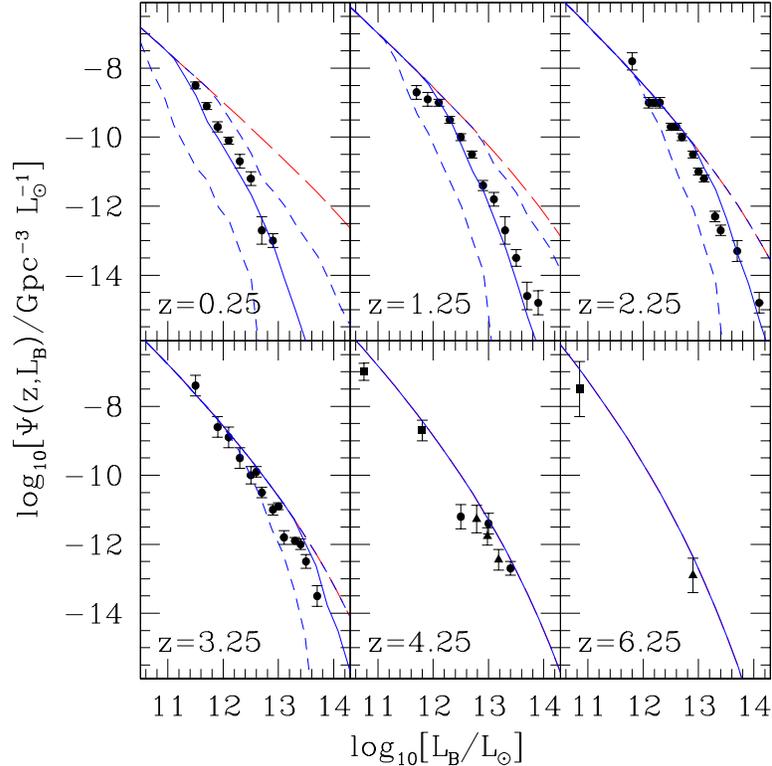,height=10.5cm}}
%\plotone{f6.eps}
\caption{Evolution of the B-band quasar luminosity function.
In all panels the long-dashed lines
are the WL03 model with $F = 1.0$,
the solid lines are our fiducial $F= 1.0$, $\epsilon_{\rm k} = 0.05$ model,
and the upper and lower short-dashed lines are models in which
$F=1.0$ and $\epsilon_{\rm k}$ is 0.025 and 0.10 respectively.  Points 
are as in Figure 1.}
\label{fig:lum2}
\end{figure}

The trends seen in Figure \ref{fig:zex} have a direct impact on the 
B-band quasar luminosity function, as shown in Figure \ref{fig:lum2}.  
From this plot it is immediately clear that the 
exclusion of $z \leq z_{\rm ex}$ objects in our fiducial models
provides an excellent fit to the data.  Less clear is just how
much of this agreement is due to our allowing ourselves a specific
choice of $\epsilon_{\rm k} = 0.05.$

We would maintain that the best answer to this question is exactly
half.  
There are two characteristic values that appear in this figure,
the first of which is the characteristic luminosity, or mass, above
which feedback effects are imprinted.  This corresponds to the
transition region in Figure \ref{fig:zex}, which we saw could be
simply shifted to higher (lower)  values by decreasing (increasing)
$\epsilon_{\rm k}$.  Yet increases in the kinetic-energy input also have the
effect of pushing  $z_{\rm ex}$ to higher redshifts.

This corresponds to the second characteristic value in Figure 
$\ref{fig:lum2}$, namely the redshift at which feedback corrections to the 
WL03 model become important.  This value is not arbitrary within
our model, but rather is fixed when $\epsilon_{\rm k}$ is chosen
to reproduce the characteristic $L_B$.  Thus the $\epsilon_{\rm k} = 0.025$
model shown in this figure not only fails to halt the formation of
a sufficient number of higher-luminosity quasars at $z=0.25$, but
fails to have any impact whatsoever at $z \geq 1.25.$  Similarly,
the $\epsilon_{\rm k} = 0.10$ model both over-estimates feedback effects at 
$z = 0.25$, and deviates from the WL03 model too soon.

\subsection{A Heuristic Aside}

To gain a heuristic understanding of our results, it is useful to
explore a Soltan-type argument (Soltan 1982; Yu \& Tremaine 2002),
relating the accretion history of black holes to their total energy
release over the history of the universe. We can simply modify the
argument to consider the total entropy injected into the IGM, 
rather than the total optical light emitted. To do so, we note
from eq. (\ref{eqn:masscrit}) that $M_{\rm b, heat} \propto E_{60}=
\epsilon_{\rm k} L_{\rm bol} t_{\rm dyn} = \epsilon_{\rm k} \epsilon_{\rm
rad} \dot{M}_{bh} c^{2} t_{\rm dyn} = \epsilon_{\rm k} \epsilon_{\rm
rad} \Delta M_{\rm bh} c^{2}$. Here, the radiative efficiency is
$\epsilon_{\rm rad}\equiv L_{\rm bol}/(\dot{M} c^{2}) \approx 0.1$
from comparing estimates of the local black hole mass density with the
quasar luminosity function from the 2dF redshift survey (Yu \& Tremaine
2002). Thus, the total mass density of gas heated above $S_{\rm crit}$
is directly proportional to the mass density in black
holes, if we ignore the time delay due to the finite bubble 
propagation speed. Specifically, we obtain
\begin{equation}
\rho_{b}(S>S_{\rm crit}) \approx 10^{4} \rho_{\rm bh} 
S_{\rm 100, crit}(z)^{-1} (1+z)^{-2} \delta_s^{-2/3} 
\left( \frac{\epsilon_{k}}{0.05} \right)
\left( \frac{\epsilon_{\rm rad}}{0.1} \right).
\label{eqn:soltan}
\end{equation}     
The comoving mass density in black holes as a function of redshift is
\begin{equation}
\rho_{\rm bh}(z)= \int_{z}^{\infty}
dz^{\prime} \int_{0}^{\infty} dL_{B} \frac{L_{\rm bol}}{\epsilon_{\rm
rad} c^{2}} \Psi(L_{B},z^{\prime}) \frac{dt}{dz'}.
\label{eqn:rho_bh}
\end{equation}
We can use eqs.\ (\ref{eqn:soltan}) and (\ref{eqn:rho_bh}) to obtain
$\Omega_{b}(S > S_{\rm crit},z)$, without recourse to Press-Schechter
theory. We can also approximately understand the exclusion redshift
$z_{\rm ex}$ from the fact that the highest density peaks are always
the first to be polluted to high entropy, just as they are also always
the first to be polluted to high metallicity.  The exclusion redshift
can therefore be approximated by:
\begin{equation}
f(S>S_{\rm crit})= \frac{\rho_{b}(S>S_{\rm crit})}{\bar{\rho_{b}}} \approx
{\rm erfc}\left( \frac{\delta_{c}}{\sqrt{2} \sigma(M,z_{\rm ex})}
\right).
\end{equation}
From this we can understand the evolution of the mass suppression
scale. Often in models of structure formation, the suppression mass
(for instance, the Jeans mass) is self-regulating and therefore traces
the non-linear mass scale (e.g., Chiu \& Ostriker 2000). This is
manifestly not the case for preheating. For instance, if preheating
affected only $\sim 2 \sigma$ perturbations, then $f(S>S_{\rm crit})
\sim 0.05$, independent of redshift. Instead, from the sharp rise in
the black hole mass density with time, we know that $f(S>S_{\rm
crit})$ increases with time. Thus, at late times, preheating affects
progressively smaller perturbations: from $\sim 6\sigma$ perturbations
at $z\sim 5$ ($M_{\rm suppress} \sim 10^{13.5} M_{\odot}$) to $\sim
1\sigma$ perturbations at $z\sim 1$ ($M_{\rm supress} \sim 10^{12}
M_{\odot}$). Once preheating affects the bulk of halos, the supply of
cold gas is cut off and the comoving luminosity density in quasars and
star formation abruptly drops.

%We can also understand why the downturn in the comoving quasar luminosity
%density (at $z \sim 2$) precedes the downturn in the comoving star
%formation rate (at $z\sim 1$). Since $L_{\rm quasar} \propto M_{\rm
%galaxy}^{5/3}$, the light in quasars is more strongly dominated by high
%mass objects than galaxies. However, high mass objects are the first
%to be affected by preheating. Thus, quasars self-terminate at a higher
%redshift than galaxies.

\subsection{Entropy and Galaxy Formation}

\begin{figure}
\centerline{\psfig{figure=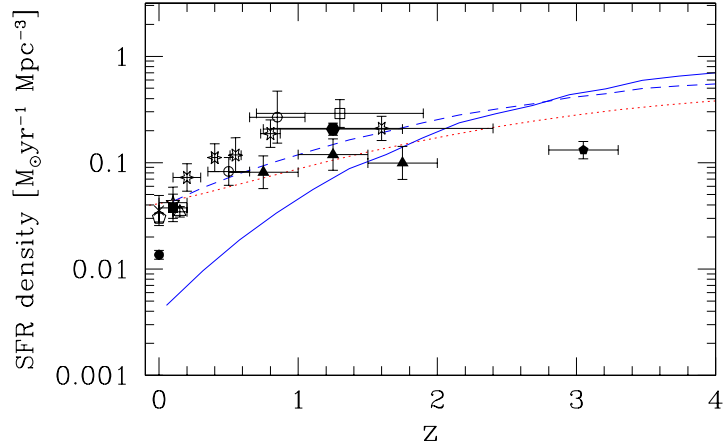,height=6.6cm}}
%\plotone{f7.eps}
\caption{Observed and predicted star formation rate densities.  
Here the solid and short-dashed  lines represent the 
$f_\star=0.1$, $f_{\star,q}=0.0$ and 
$f_\star=0.05$, $f_{\star,q}=0.003$
models, respectively, the dotted line is the fit by
Hernquist \& Springel (2003), and 
the points are taken from a wide range 
of optical, far-infrared, and 1.4 GHz measurements, 
as compiled and corrected for reddening by  Hopkins \etal (2001),
and converted to our assumed cosmology.
In particular, the symbols represent measurements by
Haarsma \etal(2000)      (six-pointed stars), 
Hopkins, Conolly, \& Szalay (2000)  (solid hexagons),
Sullivan \etal (2000)    (five-pointed stars),
Serjeant \etal (2002)    (open pentagons),
Steidel \etal (1999)     (filled pentagons),
Yan \etal (1999)         (open squares),
Treyer \etal (1998)      (filled squares),
Tresse \& Maddox (1998)  (open triangles),
Connolly \etal(1997)     (filled triangles),
Rowan-Robinson \etal(1997) (open circles),
Gallego \etal (1995)       (filled circles), and
Condon (1989)              (crosses).}
\label{fig:sfr}
\end{figure}

In Figure \ref{fig:sfr} we compare our feedback models with what has
emerged as one of the most widely-used quantifiers of galaxy
evolution, the comoving space density of the star formation rate,
$\dot \rho_\star.$   In this plot, the  barrage of multi-wavelength
observations illustrates  that  $\dot \rho_\star$   rises by an order
of magnitude between $z=0$ and  $z=1,$ and evolves much more slowly at
higher redshifts.
For reference, we also include the analytical model given by
eq.\ (2) of Hernquist and Springel (2003), with a normalization of
$\dot \rho_\star(0) = 0.04 \msun$ yr$^{-1}$ Mpc$^{-1}.$

While it is still unclear whether the evolution of $\dot \rho_\star$ 
reaches a peak around  $z \approx 1.5$
(\eg Madau \etal 1996) or stays flat to much higher 
redshifts, the primary uncertainties in this comparison are theoretical
rather than observational.  A rough estimate of the number of stars resulting
from starbursts can be taken by simply evaluating the time derivative of
the total mass contained in galaxies, scaled to the average number
of stars formed in each burst.  This gives
\be
\dot \rho_\star(z) = f_\star \, \frac{\Omega_b}{\Omega_0} \, \frac{dz}{dt} \,
\int_{M_{40}(z)}^\infty dM \, \frac{dn}{dM dz} \, 
\left[ 1 - f_S(z,M) \right],
\label{eq:fstar1}
\ee
where $f_\star$ is a ``star formation efficiency'' that parameterizes
the fraction of gas in a given starburst that is converted to stars.
Taking $f_\star$ to be a free parameter that we match to 
observations at high redshifts yields the solid line shown in this figure.  
Obviously the fall-off at lower redshifts is
 much more severe than observed, yet the reason for the discrepancy is
clear, as the majority of observed low-redshift star formation occurs 
not as bursts, but rather  though quiescent star formation in disk 
galaxies (\eg Tan, Silk, \& Balland 1999).  

This sort of star formation could simply be modeled by a term that
depends on the total (integral) mass in galaxies
\ba
\dot \rho_{\star,{\rm quiescent}}(z) &=& 
f_{\star,q} \, \frac{r_{\rm vir}}{v_c}(z) \, \frac{\Omega_b}{\Omega_0} 
\int_z^\infty dz' \times \nonumber \\
& & \, \, \, \int_{M_{40}(z')}^\infty dM \, \frac{dn}{dM dz'} \, 
\left[1 - f_S(z',M) \right],
\ea
where the dynamical time  $(r_{\rm vir}/v_c) = 30 t_{\rm dyn}$   as
given by eq.\ (\ref{eq:tdyn}) and $f_{\star,q}$ is now a parameter
that determines the average rate of quiescent star formation in disks.
Adding this contribution (with $f_{\star,q} = 0.003$) to eq.\
(\ref{eq:fstar1}) results in the dashed line shown in  Figure
\ref{fig:sfr}.  While this provides a reasonable fit  to observations,
the necessary introduction of a second free parameter makes this
comparison a bit unsatisfying.  We can conclude only  that $\dot
\rho_\star(z)$ measurements are consistent with strong-feedback
models, although they do not provide particularly strong constraints.
We note in passing, however,  that our fiducial $\epsilon_{\rm
k}=0.05$ model is consistent with observations of Lyman-break galaxies
at $z\sim3$, as only halos of $M \gtrsim 10^{13} M_{\odot}$ have 
$z_{\rm ex} \geq 3$, while Lyman-break galaxies are 
thought to reside in halos with masses $\sim 10^{12} M_{\odot}$, 
or even less (\eg Scannapieco \& Thacker 2003).

\begin{figure}
\centerline{\psfig{figure=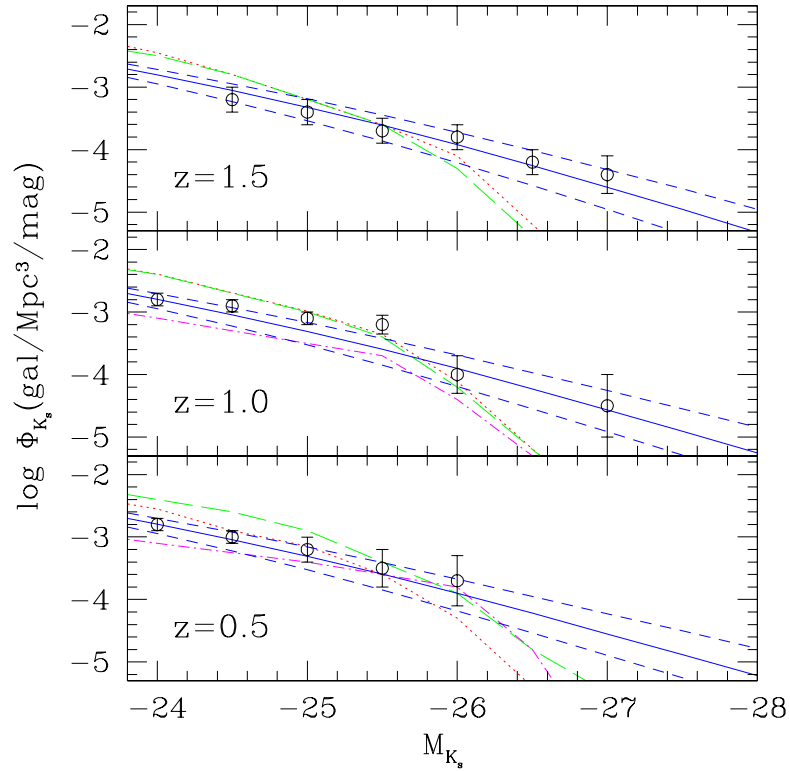,height=10.5cm}}
%\plotone{f8.eps}
\caption{The rest-frame $K_s$ band luminosity function of luminous galaxies.
In each panel the solid lines represent our fiducial $F=1$, $\epsilon_{\rm k} =0.05$
model, while the $\epsilon_{\rm k} = 0.025$ and $\epsilon_{\rm k} = 0.10$ models
are given by the upper and lower short-dashed lines respectively.
The points are the $K_s$ band observations from Pozzetti \etal (2003),
and the dotted, long-dashed, and dot-dashed lines represent
the Menci \etal (2002), Cole \etal (2000), and Kauffmann \etal
(1999) models, respectively, as compiled by Pozzetti \etal (2003).
Quasar feedback naturally turns off massive galaxy formation at low redshifts,
reproducing the observed lack off evolution in the number density of 
luminous galaxies below $z \leq 1.5$.}
\label{fig:mag}
\end{figure}

A more rigorous confrontation with observations arises from
restricting our attention to the subset of galaxies in which quasar
feedback has the strongest impact, namely those with  masses $\gtrsim
10^{12} \msun$.  Thus in Figure \ref{fig:mag} we plot the number
density of the most luminous galaxies at three representative
redshifts.  Here we restrict our  attention to $K_s$ measurements as
taken by Pozzetti \etal (2003), as such infrared measurements are most
sensitive to the overall galaxy stellar masses, rather than a
combination of stellar masses and formation histories (\eg Madau,
Pozzetti, \& Dickinson 1998).

To convert our mass function to $K$-band magnitudes in an approximate
and simple manner we make use of the Tully-Fisher relation
observed by Pierini \& Tuffs (1999), which relates the
$K'$-band absolute magnitude and circular velocities of
disk galaxies as
\be
M_{K'} = -9.66 \, v_c \, - \, 1.41.
\label{eq:PT}
\ee
For purposes of this comparison, we associate $v_c$ in this equation
with the underlying halo circular velocity, making no attempt 
to correct for galaxy morphology or differences between
$K'$ and $K_s$ bands.  This allows us to compute
the luminosity function as
\be
\Phi_{K_s}(z) = 
\int_z^\infty \, dz' \, \frac{d v_c}{d K_s} \, \frac{d M}{d v_c} 
 \frac{dn}{dM dz'} \left[ 1 - f_S(z',M) \right],
\ee
where $M$ is a function of $z'$ that is fixed by eqs.\ (\ref{eq:PT})
and (\ref{eq:vc}).  This is shown as the solid 
($\epsilon_{\rm k} = 0.05$) and short-dashed ($\epsilon_{\rm k} = 0.025$;
$\epsilon = 0.10$) lines in Figure \ref{fig:mag}.

Here we see that our simple model provides an excellent fit to the
data, so much so  that some degree of chance must be involved, given
the uncertainties.  This agreement is even more striking given the
large {\em inconsistencies} between the data  and the more detailed
galaxy-formation models compiled from Menci \etal (2002), Cole \etal
(2000), and Kauffmann \etal (1999) in this figure, particularly for the
brighter objects. Furthermore, 
this sample is by no means the only one
that illustrates a discrepancy between observations and modeling.
Rather, a large number of independent surveys are consistently
pointing to a lack of galaxy evolution below $z \lesssim 1.5.$   (\eg
Cimatti \etal 2002;
Chen \etal 2003a; Glazebrook \etal 2003; Somerville \etal 2003),
which appears to present a ``paradox'' for current  hierarchical
models (\eg Chen \etal 2003b).  Given this well-known discrepancy, is
it not perhaps the case that we obtain agreement by overlooking an
important piece of galaxy-formation physics that when accounted for
would completely change our model?

While a key issue in galaxy formation has indeed been neglected, we
would suggest that this points not to a shortcoming in our approach,
but rather to its greatest strength.    Despite their diversity,  the
literature models compiled in this plot share the same core mechanism
for arriving at the characteristic luminosity ($L^\star$), above which
the number densities of galaxies falls off precipitously.  Typically
this is associated with a post-viralization cooling condition (Rees \&
Ostriker 1977; Dekel \& Silk 1986), which  arises when adopting a
picture in which gas is shocked to the virial temperature at the
redshift of halo collapse, and then must have time to cool  into a
disk before star-formation can occur.  This can be implemented by
enforcing some variation on the condition
\be
f \,  t_{\rm cool}(M,z_{\rm col})
\geq t_{\rm obs} - t_{\rm col},
\label{eq:reesostriker}
\ee
where $t_{\rm obs}$ is time
corresponding to the redshift of observation, and
$t_{\rm cool}(M,z_{\rm col})$ is the cooling time as 
calculated from eq.\ (\ref{eq:tcool}) 
assuming a density of 180 times the mean density at $z_{\rm col}$
and a temperature equal to the viralization temperature $T_{\rm vir}
= 35 K [v_c/({\rm km} \, {\rm s}^{-1})]^{2}.$

\begin{figure}
\centerline{\psfig{figure=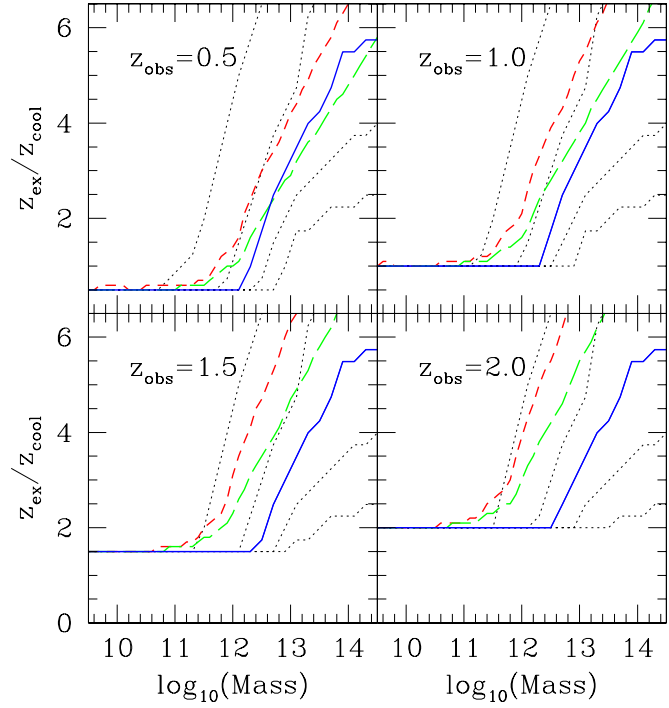,height=10cm}}
%\plotone{f9.eps}
\caption{Comparison between galaxy regulation by post-virialization
cooling and quasar feedback.  In each panel 
the solid line shows the last redshift 
of formation 
as defined by $z_{\rm ex}$ in our fiducial model, while the 
short-dashed and long-dashed lines impose post-virialization
criteria according to eq.\ (\protect\ref{eq:reesostriker}),
with $f = 1$ and $f = 1/2,$ respectively.  Finally the dotted lines
correspond (from top to bottom) to $z$ values such that 
$f_S(z,M) =$ 0.25, 0.5, 0.75, and 0.9, as in 
Figure \protect\ref{fig:zex}.}
\label{fig:zc}
\end{figure}

At the lowest redshifts, this condition provides a similar and equally
good fit to the data as our modeling, as illustrated in Figure
\ref{fig:zc}. In the upper left panel of this plot, we see that $f$
can easily be chosen to exclude star-formation in a similar subset of
halos as quasar feedback.  Yet once $f$ is fixed, the
imposition of a post-virialization cooling condition has a very
different impact on higher-redshift halos.

Although the $(1+z)^3$ increase in the mean density shortens
higher-redshift cooling times, this effect is somewhat  counteracted
by the strong $t \propto (1+z)^{3/2}$  dependence on time with
redshift.  This means that in higher-redshift objects, eq.\
(\ref{eq:reesostriker}) continues to enforce a substantial redshift
delay $(\Delta z)$  between the halo collapse and star formation
redshifts.  Thus, the reduction in objects with luminosities $\sim
L_\star$ at $z = 0$ is achieved {\em as a result} of a strong redshift
evolution in their number densities.  This generic feature, while
lessened somewhat by  more complicated modeling, is the primary source
of the discrepancy between the observations and standard models.

Furthermore, the imposition of a post-virialization cooling
criterion fails quantitatively in comparison with numerical
simulations. If no feedback or preheating is invoked, high-resolution
simulations predict that $\sim 30-40\%$ of the gas in clusters should
cool (Suginohara \& Ostriker 1998; Lewis \etal 2000; Dav\' e \etal 2001). 
This is because while $t_{\rm cool} > t_{H}$ at the virial radius, 
$t_{\rm cool} \ll t_{H}$
in the dense inner regions of clusters, resulting in catastrophic
cooling. By contrast, K-band observations indicate a cold gas fraction
of $\sim 5-10\%$ in most groups and clusters, independent of virial
temperature.  A lucid recent discussion of this overcooling problem is 
given in Balough \etal (2001).

On the other hand, our proposed feedback criterion differs from  the
customary one in that it is evaluated not at the virial radius but 
at the dense centers of halos
(which are the most highly biased and therefore most likely to have
been preheated).  Quasar feedback excludes a fixed fraction of
galaxies within halos of a given mass and formation redshift,
regardless of the redshift at which they are observed.  Thus the solid
and dotted lines in Figure \ref{fig:zc} remain fixed at all redshifts,
evolving only to enforce $z_{\rm col} \geq z_{\rm obs}$ at the
small-mass end.  This  results in the observed lack of evolution seen
in Figure \ref{fig:mag}, which naturally reproduces the observations
using only a simple transformation between $v_s$ and $M_{K_s}.$

Our point is not that a lack of $z \lesssim 1.5$ evolution can not be
accommodated in standard galaxy-formation scenarios, but rather that
this behavior can be reproduced only by including a series of detailed
corrections.  In galaxy formation models regulated by quasar feedback,
this lack of low-redshift galaxy evolution emerges naturally, imposed
by the redshift at which preheating becomes prevalent, which in turn
is fixed by the observed value of $L_{\star}$.

%corrections that enthusiasts might describe  as physical refinements
%but critics may find more reminiscent of  epicycles.  In galaxy
%formation models regulated by quasar feedback,  there is no need  for
%such debates on ``naturalness,'' rather a lack of low-redshift galaxy
%evolution is a generic feature that can not be avoided, imposed by
%the observed value of $L_\star$.

\subsection{Entropy, Clusters and the Intergalactic Medium}
\label{sect:IGM}

Clearly, the presence of quasar outflows has implications far beyond
the formation rate of supermassive black holes and the regulation of
star formation in the galaxies that surround them.   In fact, as
described above, the most-clearly observed imprint of
non-gravitational heating on cluster scales is not on galaxies
themselves, but rather on the diffuse intracluster medium (Kaiser 1991).
The implementation of strong quasar feedback as in Figure \ref{fig:Mfinal}
naturally results in an increase in ICM entropy, placing it on a higher
adiabat that prevents it from reaching a high central density during
collapse, which in-turn decreases its X-ray luminosity (\eg Tozzi \&
Norman 2001).  Furthermore for a fixed level of heating per gas
particle, this  effect is more prominent for poorer  clusters, whose
virial temperatures are comparable to this extra contribution.   As a
result, a $T^{7/2}$ relation is established in hot systems and broken
for colder systems, reconciling observations with semi-analytical studies
and numerical simulations  by establishing a fixed  entropy ``floor'' 
of $55 - 110~h^{-1/3}$~keV~cm$^{2}$.

A further constraint on the state of ICM gas comes from measurements
of the cosmic microwave background (CMB). As CMB photons are
relatively  low-energy, thermal motions in  hot ionized regions are
able to scatter them to higher energies: increasing the number of
energetic photons  and reducing the number of lower-energy ones. The
magnitude of this shift is  proportional to the Compton-$y$ parameter, the
temperature of the gas convolved with  the total column density along
a line-of-sight, and is thus most severe along sight-lines that pass
through the hot and relatively dense ICM.  This so called
Sunyaev-Zel'dovich  effect (Sunyaev \& Zel'dovich 1972) has been well
studied theoretically (\eg Aghanim \etal 1997; Refregier \etal 2000)
and detected in a number  of clusters  (\eg Grego \etal 2001; Reese
\etal 2002).
The possibility of detecting the Sunyaev-Zel'dovich effect from
galactic and quasar outflows has been explored by previous authors
(Natarajan \& Sigurdsson 1999; Yamada \etal 1999; Aghanim, Balland, \&
Silk 2000; Majumdar, Nath, \& Chiba 2001; Platania \etal 2002), and the high
optical depth measured by WMAP indicates that it might even arise from
very high redshift sources $z>10$, where Compton cooling is extremely
efficient (Oh, Cooray, \& Kamionkowski 2003).

\begin{figure}
\centerline{\psfig{figure=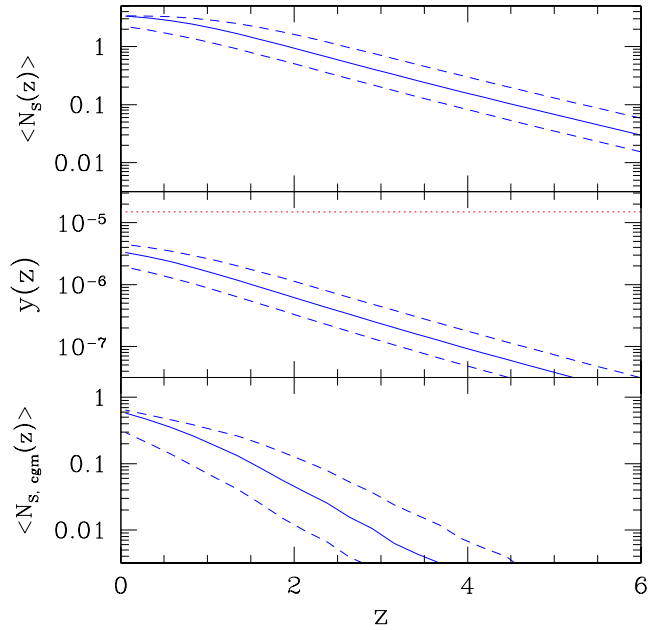,height=9cm}}
%\plotone{f10.eps}
\caption{Global impact of quasar outflows on the intergalactic medium.
{\em Top:} The mean number of $S \geq S_{\rm crit}$ outflows impacting
a random point in space as
computed from eq.\ (\protect\ref{eq:Ns0}).  Here the solid line
is given by our fiducial $\epsilon_{\rm k} = 0.05$ model, while the upper and
lower  dashed lines correspond to the $\epsilon_{\rm k} = 0.10$ and
$\epsilon_{\rm k} = 0.025$ models respectively.  {\em Center:}  Compton-$y$
parameter as computed from  eq.\ (\protect\ref{eq:y}).  Solid and
dashed lines are as in the top panel, while the dotted line is the
observational limit of $y \leq 1.5 \times 10^{-5}$ (Fixen \etal
1996). {\em Bottom:} Mass fraction of heated gas in a typical region
of circumgalactic gas, associated with the Ly$\alpha$ forest at $z
\gtrsim 2$ and the WHIM at low redshifts, as computed from
eq.\ (\protect\ref{eq:cgm}).  Lines are as in the top panel.}
\label{fig:cgm}
\end{figure}

Similarly, the $S \geq S_{\rm crit}$ regions that preceded clusters,
while more difficult to measure individually, are likely to have left
a substantial imprint on the global structure of the microwave
background.  In the upper panel of Figure \ref{fig:cgm}, we plot the
mean number of $S \geq S_{\rm crit}$  outflows impacting an arbitrary 
point in space, $\left<N_S(z)\right>$, as estimated
from  eq.\ (\ref{eq:Sfrac}).  Here we see that IGM heating is extensive
for all three models under consideration, impacting $\gtrsim 50\%$ of
the mass before the $z \lesssim 1.5$ epoch of cluster formation, even
in the lowest-energy, $\epsilon_{\rm k} = 0.025$ case.

We can estimate the overall CMB distortions from these regions as 
\ba
y(z) &=&  \frac{\sigma_T c}{m_e c^2} \int_z^{z_{\rm max}} dz'
\frac{dt}{dz'}  k T(z') n_e(z') \nonumber \\
&=& 2.4 \times 10^{-6} \int_z^{z_{\rm
max}} dz'  \frac{T_{\rm keV}(z') (1+z')^{2}}{E(z')} \left< N_S(z') \right>,
\label{eq:y}
\ea 
with $T_{\rm keV}(z) \approx \frac{S}{100 {\rm keV} \, {\rm
cm}^{2}} 8.7 \times 10^{-3} (1+z)^2 \delta^{2/3} \approx 5.2 \times
10^{-2}  \frac{S_{\rm crit}(z)}{100 {\rm keV} \, {\rm cm}^{2}}
(1+z)^2$, where we assume that mean heated regions are heated to
levels slightly higher than $S_{\rm scrit}$ and $\delta = 4$, such
that  $S \delta^{2/3} \sim 4 S_{\rm crit}(z)$.  While $y(z)$ is plotted
in the central panel of Figure \ref{fig:cgm} over the full range of
redshifts under consideration, note that only the present day value, $y(0)$,
is accessible to observation.  For all models, this value is well-below the 
observational limit on the total Compton 
distortion (Fixen 1996), but comparable to
the $\sim 10^{-6}$ estimates of the total impact from gravitationally
heated gas (\eg Refregier \etal 2000).  Thus more
detailed comparisons between future CMB 
measurements and outflow models are likely to be able to 
place strong constraints on the overall level of quasar feedback.

In the lower panel of Figure \ref{fig:cgm} we turn our attention
to even more diffuse regions of space, with overdensities of only a few.
In order to estimate the impact of quasar winds of such circumgalactic
gas, we modify eq.\ (\ref{eq:Ns}) as
\ba
\left< N_{S,{\rm cgm}}(z) \right>
&=&  \frac{1}{\bar \rho_{b,0}} \int_z^\infty dz' 
\int^\infty_{M_{\rm 40}(z')} d M  \times  \\
& & \qquad 
\, M_{\rm b, min} \{z',\delta[z',z_{\rm cgm}(z)],z,M \}  \times \nonumber \\
& & \qquad b[z',M,z_{\rm cgm}(z),M_{40}(z)]  
\, \frac{d L_B}{d M} \Psi[z,L_B(z,M)], \nonumber
\label{eq:cgm}
\ea 
where now $z_{\rm cgm}(z) < z$ is chosen such that at a redshift
$z$ the region under consideration has ceased to expand, but has not
yet collapsed into a bound structure, which can be achieved by
requiring  a ``turn-around overdensity'' associated with a collapse
parameter $\theta = \pi$ in eqs.\ (\ref{eq:theta1}) and
(\ref{eq:theta2}).  Finally we take $M_{\rm 40}(z)$ as a rough estimate
of the  the typical Jeans mass in the post-reionization IGM (Barkana
\& Loeb 2001).

Surprisingly, only about $5 \%$ of the circumgalactic gas has been
impacted by outflows at $z \sim 2$, but the fraction of heated gas
increases to $\sim 30-40\%$ at $z\sim 0$. This may seem surprising,
since the quasar population peaks at $z\sim 2$, after which the
comoving density of sources falls sharply. The reason is the finite
travel time of the bubbles, as given by equation
(\ref{eqn:texpand}). Thus, while the winds affect the sites of massive
galaxy and quasar formation (which are highly clustered) early on,
they only impact the low density IGM at late times.  In particular,
while quasar feedback has a large impact on the collapse of non-linear
structures, its impact on the $\delta \sim 5$, $z \gtrsim 2$ regions
associated with the Lyman-$\alpha$ forest is expected to be  small.
The feedback assumed in our modeling is therefore completely
consistent with the observed properties of this material, although it
remains that specialized analyses may be  able to detect (or rule-out)
its presence.

At lower redshifts, the circumgalactic gas acquires temperatures that
tend to be described as ``warm-hot.''  In fact,  numerical simulations
have shown that by $z= 0$ a large fraction ($\sim 30-40 \%$) of
baryons have been gravitationally heated to temperature of order $10^5
- 10^7$ K, and that this warm-hot intergalactic medium (WHIM) extends
into areas  with overdensities of only a few (\eg Cen \& Ostriker
1999; Dav\' e \etal 2001).  At $z=0$ Dav\'e \etal (2001)   find the
temperature of this gas to be $\sim 10^{5.5}$ in $\delta \sim 5$
regions (see also Nath \& Silk 2001).   Taking $S_{\rm crit} = 150$
keV and $\delta \sim 5$  in the $50\%$ of the circumgalactic gas
impacted by quasar outflows by $z \sim 0$ results in a similar value.
Thus quasar feedback represents an energy source that is comparable to
gravitational heating in the WHIM and is likely to have a large impact
on its properties. Indeed, if production of the WHIM via quasar
outflows is important, then the majority of the IGM will not be
graviationally shocked at late times, but instead evolve
isentropically. While observational constraints on this material are
relatively weak at the moment, as in the CMB case, future studies are
likely to provide a sensitive  probe into heating of this mutiphase
gas  (\eg Shull, Tumlinson, \&  Giroux 2003) and the impact of quasar
feedback.

\subsection{Relation to Baryonic Stripping}

%While the energy input from
%quasars is likely to scale as the depth of the gravitational potential
%wells in which they are contained $E \propto M_{\rm bh} \propto M_{\rm
%halo}^{5/3}$, the energy of supernova (SN) driven winds scales
%approximately as the total mass in stars formed $E \propto M_\star \propto
%M_{\rm gas} \propto M_{\rm halo}.$ Thus starburst outflows are most
%efficient in escaping from dwarf galaxies,  whose formation at high
%redshift ushered in the first period of intense  star formation in the
%universe.

\begin{figure}
\centerline{\psfig{figure=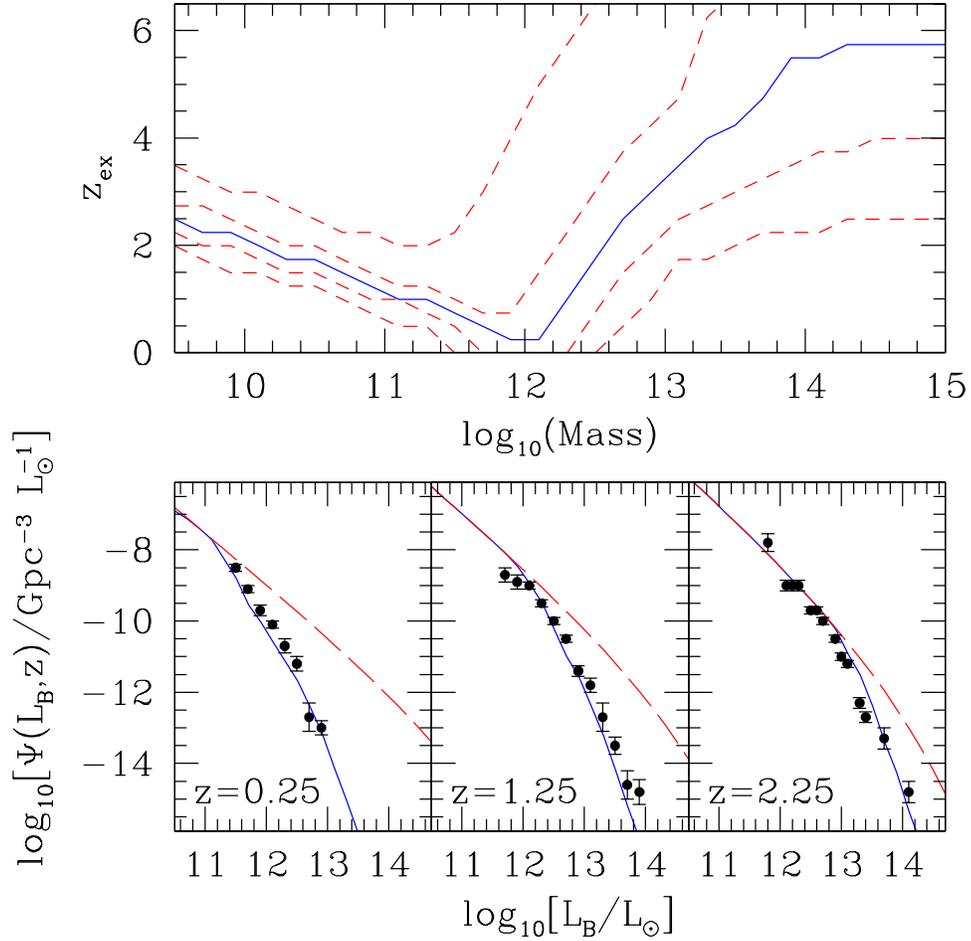,height=14cm}}
%\plotone{f11.eps}
\caption{Comparison between feedback through $S \ge S_{\rm crit}$ heating 
and the baryonic stripping of objects, which dominates in high-redshift
starbursts.  {\em Top:} As in Figure \protect\ref{fig:zex}, 
the solid line shows $z_{\rm ex}$ such that $\left< 
N_S(z_{\rm ex},M) \right> = 1,$ while the short-dashed lines correspond
(from top to bottom) to $z$ values such that $f_S(z,M) = 1 - \exp \left[
- \left< N_S(z,M) \right> \right] =$ 0.25, 0.5, 0.75, and 0.9.
While quasars are able to strip the gas out of small perturbations
that are not heated above $S_{\rm crit},$ this occurs relatively late, 
long after the majority of objects in this mass range have formed.
{\em Bottom:} Quasar luminosity function.  In all panels the solid line  
is our fiducial model, but including the additional feedback from 
baryonic stripping, while the long-dashed lines are the WL03 model.
Points are in Figure 1.  Stripping occurs too late to have any impact on
these results.}
\label{fig:strip}
\end{figure}

Finally, we relate the entropy-driven process of quasar feedback to
the effects of starburst-driven winds. While previous work (Scannapieco, Ferrara, \& Broadhurst 2000; Thacker, 
Scannapieco \& Davis 2002) has shown that dwarf outflows are relatively
inefficient at heating gas to $S \geq S_{\rm scrit}$ levels, they
are nevertheless able to impede the formation of galaxies with masses 
$\sim 10^9 M_\odot$, through the transfer of momentum.
In this case an SN driven shock accelerates the baryons in a neighboring
perturbation to above their escape velocity, stripping them entirely
from their associated dark matter and halting their collapse.
Setting the mass of the perturbation times its escape velocity
equal to the impinging momentum gives a rough estimate as to 
when such ``baryonic stripping'' will take place:
\be
M_{\rm b,p} \, v_{\rm p} \leq \left( M_{\rm s} \, v_{\rm s} \right ) \left(
\frac{\pi R_{\rm p}^2}{4 \pi R^2_{\rm s}} \right),
\ee
where $M_{\rm b,p}$, and $R_{\rm p}$ are the baryonic
mass and radius of the perturbation, while 
$v_p = \sqrt{G M_{\rm p}/(2 R_{\rm p})}$, with $M_p$ the total mass
of the perturbation.  Relating $M_s$ and $R_s$ to $E_{\rm 60}$ through
eq.\ (\ref{eq:RST}) and solving for the (physical) radius within which
baryons will be stripped yields
\be
R_{\rm strip} = 35 \, {\rm Mpc} \, E_{60} \, 
             \frac{\delta_s}{\delta _p^{5/3}} \, M_{12}^{-4/3} \, (1+z)^{-2},
\ee
where $\delta_s$ and $\delta_p$ are the average overdensity of the
expanding shock and the perturbation respectively, and $M_{12}$ is the
mass of the perturbation in units of $10^{12} \msun.$  
We can therefore include baryonic stripping in our modeling of quasar 
feedback by 
rewriting eq.\ (\ref{eq:Ns}) to account for the interactions
in which $R_{\rm strip} \geq R \geq R_{\rm heat}$.  This gives
\ba
M_{\rm b,strip}(z',\delta_s,z,M) & \equiv  {\rm min} \left\{ 
{\rm max} \left[ M_{\rm b,heat}(1,z',M) \, , \right. \right. 
\qquad \qquad \,\\
& \, \, \left. \left.
 \frac{4 \pi}{3} {\bar \rho_b} \, \delta_s \, R_{\rm strip}^3(z',M) \right]
\, ,\,\frac{4 \pi}{3} {\bar \rho_b} \, \delta_s \, R_s^3(z,z',M) \right\},
\nonumber
\ea
where we estimate $\delta_p/\delta_s \approx 3$.  The resulting
$z_{\rm ex}$  curves are shown in the upper panel of Figure
\ref{fig:strip}  for our fiducial  ($F =1$, $\epsilon_{\rm k} = 0.05$)
model.  Here we see that, due to the strong clustering and high masses
of most quasar sources,  baryonic stripping has an impact only on
late-forming low-mass halos, while  $S \geq S_{\rm crit}$ heating
dominates for objects with masses  $\gtrsim 10^{12} \msun$.
Furthermore, such events occur well after the typical formation
redshift of these small halos, and thus have a negligible impact on
the quasar luminosity function as shown in the lower panels of this
figure.  Thus lower-redshift quasar feedback is
completely different in nature from high-redshift SN driven-starbursts,
and is driven almost exclusively by $S \geq S_{\rm crit}$ heating.

\section{Discussion}

In this work, we have considered the impact of quasar outflows on
cosmological structure formation on all scales.  By associating only a
tiny fraction of a quasar's bolometric energy with an outflow of the form
observed in radio jets or broad-absorption-line winds, we have shown
that such objects can drastically transform our picture of the $z
\lesssim 2$ universe, resolving a number of long-standing issues.

Quasar feedback ties together the densest objects at the center of
galaxies with the most diffuse regions of intergalactic gas,
impacting all intermediate structures much as SN explosions impact
all environments within the interstellar medium.  As
outflows heat large regions of the IGM to entropies above $S_{\rm
crit}$, they regulate their own formation, resulting in the observed
fall-off in the number density of quasars below $z \approx 2$.  On larger
scales, quasar feedback halts the formation of $M \gtrsim 10^{12}
\msun$ galaxies, providing an alternative mechanism for regulating galaxy
formation than the usually-imposed condition of post-virialization
cooling.  Furthermore, while $L_\star$ is fixed in a standard
picture only as a result of strong redshift evolution, this value
appears in a feedback model as the nonlinear mass scale at the
redshift at which quasar outflows heat the denser regions of gas to
$S \geq S_{\rm crit}$ values.   Thus the observed lack of $z \lesssim
1.5 $ galaxy evolution, while arising only though detailed corrections
in a standard picture, is a general and unavoidable feature of a model
regulated by quasar feedback.

On intergalactic scales, feedback heats the denser regions of gas
to values just above $S_{\rm crit}$, providing a natural explanation
for why the ICM of clusters of all sizes has been generally preheated
to entropy values {\em just above} the minimum excess observable.
Furthermore, while the strong clustering of quasars precludes them from
having a serious impact on the somewhat overdense $z \gtrsim 2$ regions
associated with the Lyman-$\alpha$ forest, their global impact
may provide us with sensitive observational probes through Compton 
distortions  in the CMB and the multiphase structure of the ($z = 0$) 
WHIM.

While in a self-similar, hierarchical model, preheating by progenitors
would be independent of the final host halo mass scale, there are
three major features of quasar outflows  that break this
self-similarity.  The first is that energy injection is a non-linear
function of halo mass, $E \propto M_{\rm bh} \propto M_{\rm
halo}^{5/3},$ which causes the largest objects to be most impacted.
Secondly, the presence of a critical entropy imposes a scale into the
problem, as the number of supercritical outflows is self limiting,
and thus does not greatly exceed one even for the largest objects.
Finally, high mass objects are more strongly clustered, and therefore
more likely to collapse out of preheated gas than low mass objects.  A
useful analogy is that of metallicity: the most overdense regions
will tend to be polluted to high metallicity and high entropy first.

The three-phase (McKee \& Ostriker 1977) model so  much revolutionized
our understanding of the interstellar medium that it naturally lead to
widespread interest in galaxy formation though intergalactic shocking
from starbursts (\eg Ostriker \& Cowie 1981) and even ``quasar
explosions'' (Ikeuchi 1981).  While we now believe structure formation
is driven instead by dark-matter gravitational collapse, is
there not a role for outflows in structure {\em regulation}?

Today, evidence for feedback effects from high-redshift starburst
galaxies is better described as overwhelming than mounting.  Analyses
of \CIV lines in quasar absorption spectra uncover an IGM that has been
widely and inhomogenously enriched with stellar material (\eg Cowie \etal
1995; Rauch, Haehnelt, \& Steinmetz 1997) out to very high redshifts
(Songaila 2001) and  turbulently ``stirred''  at $z \sim 3$ (Rauch,
Wallace, \& Barlow 2001).  Large numbers of outflowing starbursts have
been directly detected,  both in optical and infrared observations at
$3 \lesssim z \lesssim 4$ (Pettini \etal 2001) and in optical
observations of lensed galaxies at  $4 \lesssim z \lesssim 6.5$ (Frye,
Broadhurst, \& Benitez 2002; Hu \etal 2002).  Furthermore it is clear
from both analytical and numerical studies that these winds have an
important effect on the gas in growing  pre-virialized density
perturbations, accelerating their baryons above the escape velocity
and thereby regulating the formation of further such dwarves.

At the same time, evidence of similar feedback effects at low redshift
is staring us in the face.  At $z \sim 2$ a large number of
highly-clustered quasars are observed to be pummeling the denser
regions of space, providing a natural energy source for the
substantial preheating detected in X-ray observations of galaxy groups
and clusters at $z \sim 0$.  These {\em are} the collapsed gaseous
halos that have failed to form into galaxies, and they look nothing
like the post-virialization clouds posited in standard  scenarios.
Can we really expect non-gravitational heating to be the primary
driver of their properties and yet have played only a secondary role
in the formation history of large galaxies?

As surveys scour the sky, reconstructing the history of galaxy formation
in our universe, they only serve to highlight the interdependence between 
these star-forming regions and the diffuse gas that surrounds them.
Similarly, a large range of numerical studies have suggested that
there is more to the thermal and accretion history of the IGM than we
had first imagined.  By complementing galaxy surveys with detailed
measurements of Compton distortions in the CMB, probes into the 
structure of the low-redshift WHIM, and perhaps even specialized analyses
of the Lyman-$\alpha$ forest, this history is accessible to us.
Far from an abstract conjecture, quasar-regulated structure formation
provides a clear and testable alternative to current low-redshift
modeling, a postulated missing link that may soon be unearthed by
multi-wavelength excavations into our cosmic past.

\acknowledgments 

We are grateful to Arif Babul, Emanuele Daddi, Andrew King, Hui Li,
Jaron Kurk, Crystal Martin, Daniel Proga,   Volker Springel, Rob
Thacker,  Dmitri Uzdensky, Mark Voit, and  J.\ Stuart Wyithe, as well
as an anonymous referee for helpful comments and suggestions.  This
work was supported by the National Science Foundation under grant
PHY99-07949.

\fontsize{10}{10pt}\selectfont

\newpage

\begin{figure}
\centerline{\psfig{figure=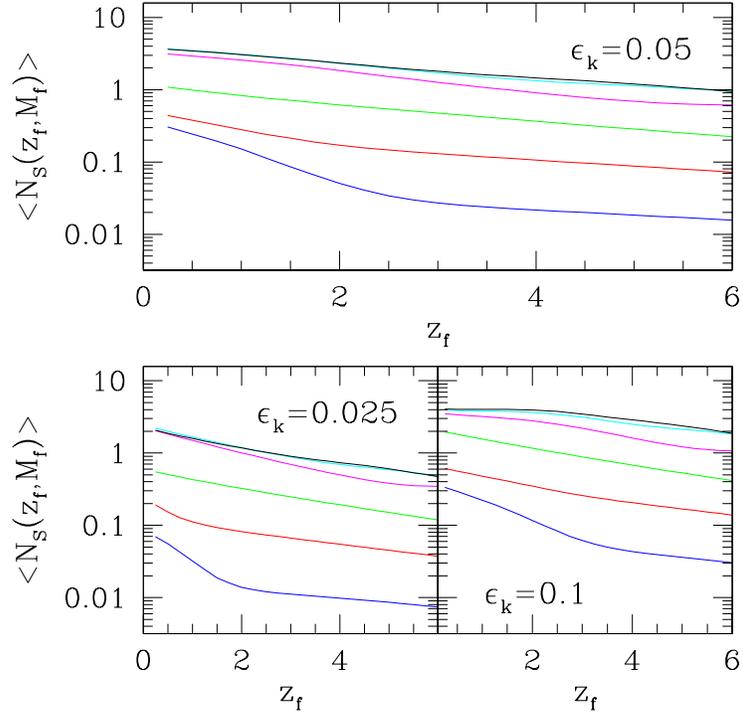,height=10cm}}
%\plotone{f3.eps}
\caption{Average number of $S \geq S_{\rm crit}$ quasar outflows
impacting objects with various formation masses and redshifts.
Panels are labeled  
by their $\epsilon_{\rm k}$ values, and in all cases the curves represent
masses (from top to bottom) of $10^{15} \msun$, $10^{14} \msun$,
$10^{13} \msun$, $10^{12} \msun$, $10^{11} \msun$, and $10^{10} \msun$.}
\label{fig:Mfinal}
\end{figure}

\end{document}